\documentclass[USenglish,oneside,twocolumn]{article}

\usepackage[utf8]{inputenc}
\usepackage[big]{dgruyter_NEW}
\usepackage{amsmath}
\usepackage{amssymb}
\usepackage[hyphens]{url}
\usepackage{hyperref}
\usepackage[linesnumbered, ruled, vlined]{algorithm2e}
\usepackage{amsmath}
\usepackage{comment}

\usepackage{array}
\newcolumntype{L}[1]{>{\raggedright\let\newline\\\arraybackslash\hspace{0pt}}m{#1}}
\newcolumntype{$}{>{\global\let\currentrowstyle\relax}}
\newcolumntype{^}{>{\currentrowstyle}}

\begin{document}

\author*[1]{Noah Apthorpe}

\author[2]{Danny Yuxing Huang}

\author[3]{Dillon Reisman}

\author[4]{Arvind Narayanan}
  
\author[5]{Nick Feamster}

\affil[1]{Princeton University, Department of Computer Science, E-mail: apthorpe@cs.princeton.edu}
  
\affil[2]{Princeton University, Department of Computer Science, E-mail: yuxingh@cs.princeton.edu}
    
\affil[3]{Princeton University, Department of Computer Science, E-mail: dreisman@princeton.edu}
      
\affil[4]{Princeton University, Department of Computer Science, E-mail: arvindn@cs.princeton.edu}
        
\affil[5]{Princeton University, Department of Computer Science, E-mail: feamster@cs.princeton.edu}

\title{\huge Keeping the Smart Home Private with Smart(er) IoT Traffic Shaping}

\runningtitle{Keeping the Smart Home Private with Smart(er) IoT Traffic Shaping}

\begin{abstract}
{The proliferation of smart home Internet of things (IoT) devices presents
unprecedented challenges for preserving
privacy within the home.
In this paper, we demonstrate that a passive network observer (e.g., an
Internet service provider) can infer
private in-home activities by analyzing Internet traffic from commercially
available smart home devices \textit{even when the devices use end-to-end
transport-layer encryption}.
We evaluate common approaches for defending against these types of traffic analysis
attacks, including firewalls, virtual private
networks,
and independent link padding,
and find that none sufficiently conceal user activities
with reasonable data overhead.
We develop a new defense, ``stochastic traffic padding'' (STP), that
makes it difficult for a
passive network adversary to reliably distinguish genuine user activities from generated
traffic patterns designed to look like user interactions.
Our analysis provides a theoretical bound on an adversary's ability to accurately
detect genuine user activities as a function of the amount of additional cover traffic
generated by the defense technique.}
\end{abstract}

\startpage{1}


\maketitle

\vspace{-46pt}
\section{Introduction}
\vspace{-8pt}
\label{sec:intro}

Internet-connected consumer devices (``Internet of things,''  ``IoT,'' or ``smart home''
devices) have rapidly increased in popularity and availability over the past several
years.
Many smart home devices have always-on sensors that (1)~capture users' offline
activities in their living spaces and (2)~transmit information about these
activities outside of the home, typically to cloud services run by device
manufacturers. 
Such communication introduces privacy concerns, not only because of the 
data that these devices collect and send to third parties, but also because {\em
the very existence of traffic at all} can reveal sensitive and private information
about the activities of a home's occupants.

\begin{figure}[t]
\centering
\includegraphics[width=0.49\textwidth]{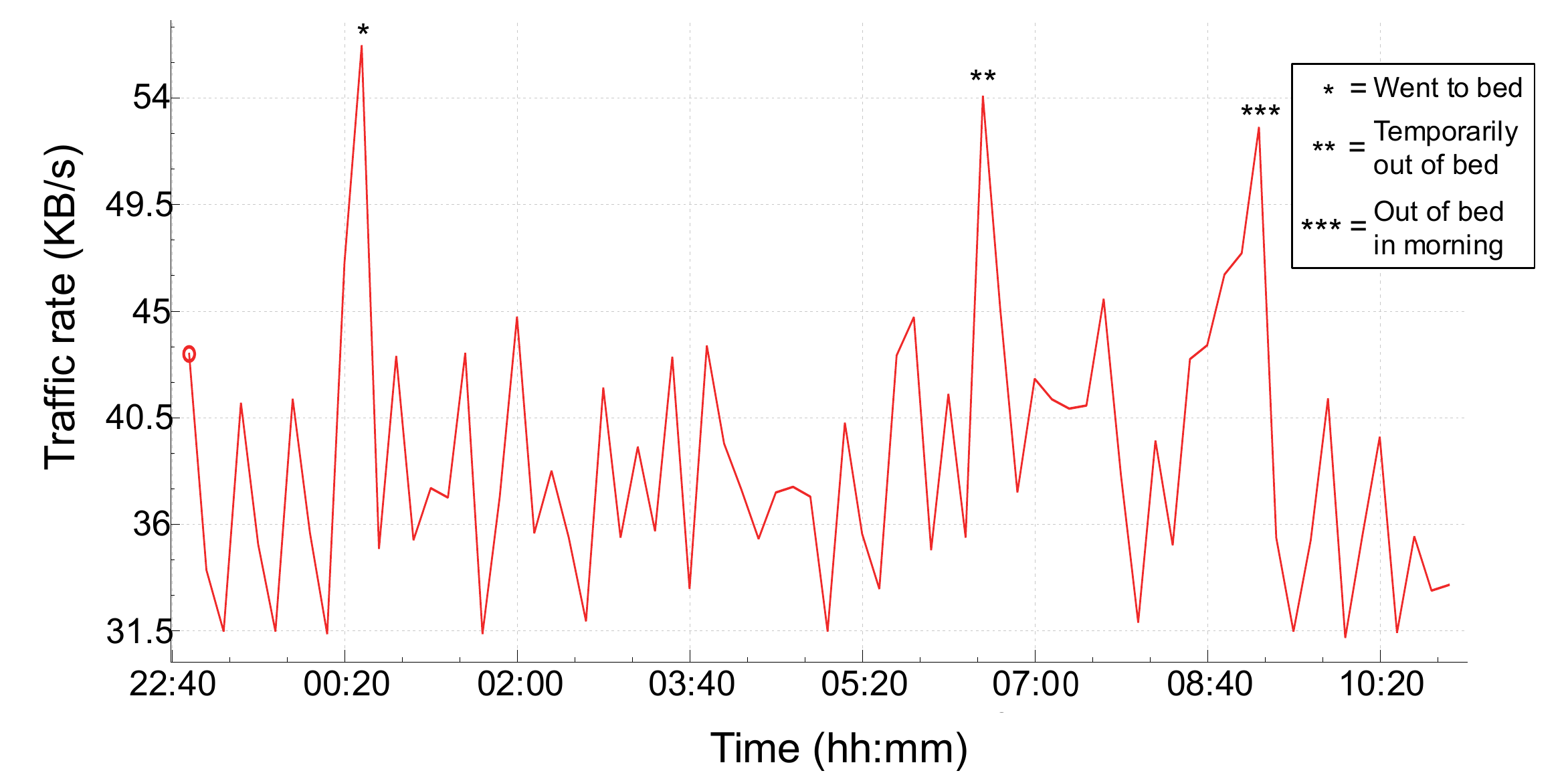}
\caption{Traffic rate to and from a Sense sleep monitor over a 12-hour period. User activities are clearly visible as traffic spikes. }
\label{fig:sleep-pcap}
\end{figure}

In this paper, we demonstrate that despite broad adoption of transport layer
encryption, smart home traffic \emph{metadata}---specifically, 
traffic volumes---is sufficient for a passive
network adversary to infer users' sensitive in-home activities
(Figure~\ref{fig:sleep-pcap}). 
As with phone metadata~\cite{mayer2016evaluating}, it is possible to
learn a great deal about devices and users from Internet metadata alone.   We
present an attack on user privacy using metadata from smart home devices that
is effective \textbf{even when devices use encryption}
(Section~\ref{sec:attack}). The attack involves 
inferring times and types of user
activities from device traffic patterns.

We demonstrate this attack on commercially available smart home
devices. For example, traffic rates from a Sense sleep
monitor reveal consumer sleep patterns, traffic rates from a Belkin Wemo
switch reveal when a physical appliance in a smart home is turned on or off,
and traffic rates from a Nest Cam Indoor security camera reveal when a user
is actively monitoring the camera feed or when the camera detects motion in a
user's home.

The effectiveness of this attack across smart home device types and
manufacturers motivates the development of general, easy-to-deploy techniques for protecting
user privacy in smart homes. 
Conventional approaches, such as firewalls, virtual private networks (VPN), and independent link padding (ILP), may break device functionality, allow user activity inference in certain scenarios, or unacceptably increase data usage (Section~\ref{sec:solution}).
Although any Internet metadata protection technique will involve
some overhead, there is a need for low-cost tunable methods that allow users to
trade off how much they are willing to spend for a guaranteed degree of
privacy. 

We therefore present a new traffic shaping algorithm, ``stochastic traffic padding (STP)'' (Section~\ref{sec:iilp}).
STP performs traffic shaping during user activities and selectively injects cover traffic during other
time periods. This makes it difficult for an adversary to distinguish traffic corresponding to 
genuine user activities from cover traffic mimicking user activities.
With STP, an adversary's confidence in detecting a genuine user activity
is inversely proportional to the bandwidth overhead that results from the injected
traffic.
We demonstrate this relationship both in theory and empirically by performing
STP on traffic traces from real smart home devices.

We also present an implementation of STP that can run on Linux-based network
middleboxes, including smart home hubs, Wi-Fi access points, and home gateway routers (Section~\ref{sec:implementation}).
This implementation demonstrates that a small amount of additional traffic
padding can significantly
reduce adversary confidence, suggesting that STP may be practical for a wide range
of deployment cases, including as a module that could run on a home network's gateway
router.

In summary, this work makes the following contributions:
\begin{enumerate}
	\itemsep=-1pt
\item We demonstrate that in-home user activities can be inferred from 
smart home Internet traffic volumes alone, even when the traffic is protected with end-to-end
encryption.
\item We evaluate conventional defenses against traffic analysis attacks in terms
of privacy protection, network delay, and traffic overhead.
\item We present stochastic traffic padding (STP), which provides
tunable protection against user activity inference 
with considerably less bandwidth overhead
than existing approaches.
\item We analyze the performance of STP using traffic traces from real
devices and develop an implementation that could be used in real smart homes.
\end{enumerate}
\vspace{-24pt}
\section{Threat Model}
\vspace{-4pt}
\label{sec:threat}

We are concerned with the ability of a passive network observer to infer users' in-home activities from smart home Internet traffic metadata. Traffic rates, network protocols, source and destination addresses, interpacket intervals, and packet sizes are accessible to many entities.
These potential adversaries may be incentivized to discover user behaviors, in opposition to the preferences of privacy-conscious device owners. 
We divide our threat model into two distinct classes of adversaries with differing visibility into the home network:

\textbf{Local adversaries.}
Local adversaries are entities that can view traffic within the smart home local area network (LAN). Example local adversaries include malicious smart home devices, compromised or ISP-controlled home routers, and Wi-Fi eavesdroppers such as neighbors and  wardrivers.
All local adversaries can view MAC addresses, send times, and sizes of Wi-Fi packets.  Local adversaries that are unable to associate with the smart home Wi-Fi network (e.g., neighbors without the WPA2 key) cannot access other packet information.  Local adversaries associated with the smart home network (e.g., malicious devices) can also view IP headers and transport layer headers of all packets as well as the contents of non-encrypted DNS packets.

\textbf{External adversaries.}
External adversaries are entities that can view smart home traffic only after it has left the home LAN. External adversaries can obtain the times, sizes, IP headers, and transport layer headers of all packets leaving the smart home gateway router. External adversaries cannot view local Wi-Fi traffic within the home and therefore cannot access device MAC addresses for identification purposes. Since most home gateway routers act as network address translators (NAT), we assume that all smart home traffic obtained by an external adversary has had source IP addresses rewritten to the single public IP of the smart home. Example external adversaries include ISPs, government intelligence agencies, and other on-path network observers.

\textbf{Restrictions on both adversaries.}
We assume that packet contents are encrypted and inaccessible to the adversary. In fact, most smart home devices we tested use TLS/SSL when communicating with cloud servers. Given the increasing focus on security in the IoT community, encrypted communications will likely become standard for smart home devices. 
By ignoring traffic contents, our user activity inference attack (Section~\ref{sec:attack}) indicates that sensitive information about user behaviors is still at risk even when industry best practices for data encryption are in place.  

We assume that the adversary can obtain and independently evaluate smart home devices. This allows the adversary to observe device traffic patterns under various use cases.  The adversary can also continuously monitor smart home traffic if it improves activity inferences. However, the adversary is not active and does not manipulate traffic to or from the smart home. 

\textbf{Adversary prior knowledge.}
We assume that the adversary may have 
\textit{a priori} knowledge about the characteristics of some kinds of in-home activities (e.g., when people generally leave for work or go to bed)
beyond what can be observed from network traffic. 
Adversary prior knowledge limits both the usefulness of traffic rate analysis and the maximum effectiveness of traffic shaping defense techniques \textit{for particular adversaries}. For example, a general adversary may know that it is more likely for a smart home occupant to go to work in the morning than in the evening. This means that smart home device traffic patterns indicating that a home occupant leaves for work in the morning are, all else being equal, more likely to be genuine than those in the evening. However, if a specific adversary, such as a neighbor, physically sees a smart home occupant  leaving for work in the evening, no amount of traffic analysis or defensive traffic shaping will affect this knowledge. Nevertheless, defensive traffic shaping may still prevent other adversaries without this prior knowledge from inferring when the occupant leaves for~work.

\textbf{Network delays.}
Some external adversaries may not see smart home traffic until it has traveled several hops from the home gateway router. Network delays due to congestion or other quality of service (QoS) queueing could therefore have perturbed packet timings.  We disregard these perturbations because they will be insignificant relative to the timescale of user activities.

\vspace{-18pt}
\section{Experiment Setup} 
\vspace{-6pt}
\label{sec:setup}

We set up a laboratory smart home environment with several commercially available IoT devices as a testbed for performing our activity inference attack and evaluating privacy protection strategies.

We configured a Raspberry Pi 3 Model B as an 802.11n wireless access point.\footnote{We followed these instructions to set up the Raspberry Pi Wi-Fi access point: \url{https://www.raspberrypi.org/documentation/configuration/wireless/access-point.md}}
The Raspberry Pi configuration code is publicly available for research use at \url{https://github.com/NoahApthorpe/iot-inspector}.
Wi-Fi compatible smart home devices were connected directly to the Raspberry Pi Wi-Fi network.  The remaining devices were connected via Bluetooth to an Android smartphone running the latest version of the devices' corresponding mobile applications. This smartphone was itself connected to the Raspberry Pi Wi-Fi network.
This setup allowed us to record all network traffic to and from the smart home devices.
We recorded traffic from the Wi-Fi interface of the Raspberry Pi to model a local adversary and from the wired interface to model an external adversary.

The commercially-available IoT devices included in our laboratory smart home are listed in Appendix Table~\ref{fig:dns-queries}.
These devices are by no means exhaustive of the wide range of available IoT smart home products.  
However, they encompass a variety of device types, manufacturers, and privacy concerns. 
Given the effectiveness of user activity inference on traffic from all tested devices (Section~\ref{sec:attack}), we believe that smart home owners should be concerned about traffic rate metadata across all types of smart home products.  
\vspace{-18pt}
\section{User Activity Inference Attack}
\vspace{-6pt}
\label{sec:attack}

We present an attack by which a passive network observer can infer in-home user behaviors from smart home device Internet traffic metadata. The attack is applicable to most currently available smart home devices and will remain an issue for new  devices released in upcoming years. Without changes in developer practices or adoption of privacy protection techniques, smart home occupants will risk network observers learning about their in-home activities. 
Early versions of this attack~\cite{apthorpe2017smart, apthorpe2017closing} have received press attention~\cite{chirgwin, coldewey, stark} for their relevance to modern IoT home devices. 

\vspace{-12pt}
\subsection{General Attack Technique}
\label{sec:general-attack}
\vspace{-6pt}
As a preliminary step, the adversary identifies devices in the smart home and separates traffic metadata by device.
Banner grabbing~\cite{antonakakis2017understanding, durumeric2015search, fachkha2017internet}, fingerprinting~\cite{shamsi2017faulds, shamsi2014hershel, caballero2007fig}, and acquisitional rule-based engines~\cite{feng2018acquisitional}, have all been proposed as methods of operating system and device identification.
Building on these methods, we note that
external adversaries and local adversaries with network access can use DNS queries and destination IP addresses to uniquely fingerprint smart home devices, even when the devices are behind a NAT (Section~\ref{sec:fingerprinting-demux}). 
Local adversaries without network access can use the first three bytes of device MAC addresses (the organizational unique identifier) to identify device manufacturers, followed by specific device identification based on traffic rate characteristics~\cite{apthorpe2017closing}.

The adversary then observes changes in traffic rates to determine the timing of user activities. Most user interactions with smart home devices occur at discrete time points or over short periods surrounded by extended periods of no interaction. 
For example, turning on a lightbulb, falling asleep, querying a personal assistant, and measuring blood pressure do not occur continuously. 
User activities generally cause noticeable changes in device traffic rates while or shortly after they occur. These changes can be brief ``spikes,'' longer ``hills'', or sometimes ``depressions'' visible on bandwidth graphs and detectable by automated threshold methods based on standard deviations from mean traffic rates or more sophisticated machine learning methods. 

Finally, the adversary infers the types of activities that cause observed traffic rate changes.
The limited-purpose nature of most smart home IoT devices makes this possible.
Users interact with traditional computing devices, such as PCs and smartphones, for a variety of purposes, making it difficult to associate any particular change in network traffic rate with a specific activity. 
In comparison, once an attacker has identified the identity of a particular smart home device, it is often trivial to associate specific traffic rate changes with user activities. 
For example, smart outlets generally have only two functions (turning on and off) and send very little background traffic. 
Spikes in traffic rate at a particular time therefore clearly imply that the outlet was turned on or off at that time. 
Many IoT devices also exhibit very regular traffic patterns, making it easy to correlate user interactions of particular types with traffic patterns from test devices operated by an adversary. 

\vspace{-12pt}
\subsection{Attack on Real Devices}
\vspace{-6pt}
\label{sec:attack-on-real-devices}
We have demonstrated the effectiveness of user activity inference on fourteen commercially available smart home devices as follows. 

\vspace{-12pt}
\subsubsection{Device Fingerprinting \& Traffic Demultiplexing}
\begin{figure*}[t]
\centering
\includegraphics[width=0.98\textwidth]{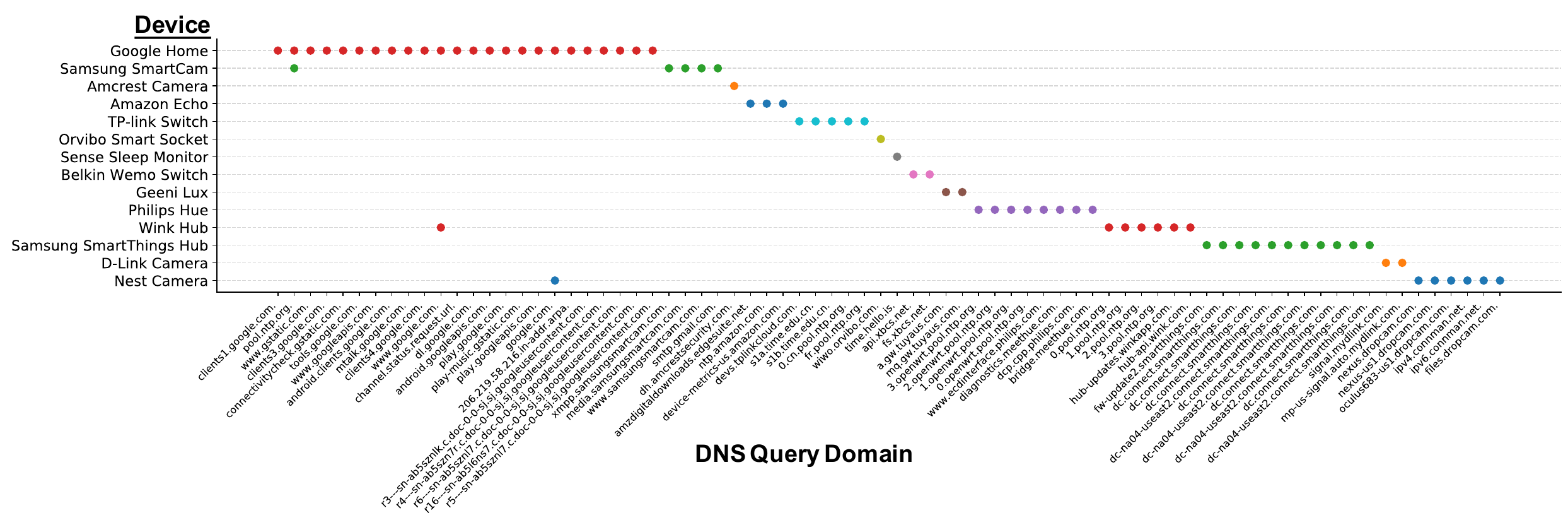}
\caption{All tested IoT devices send DNS requests for unique and mostly non-overlapping sets of domains. This allows the destination IP addresses of packets from these devices to serve as fingerprints for device identification. \vspace{-12pt}}
\label{fig:dns-fingerprint}
\end{figure*}

\vspace{-6pt}
\label{sec:fingerprinting-demux}
External adversaries can separate individual traffic flows and assign them to specific devices based on destination IP addresses, even if source IP addresses and ports are rewritten by a smart home gateway NAT.
All devices we tested send traffic to unique and mostly nonoverlapping sets of destination IP addresses and make corresponding sets of DNS requests (Appendix Figure~\ref{fig:dns-fingerprint}).  Additionally, many devices queried individual domains that uniquely identify the device or manufacturer without requiring the use of a set-based fingerprint (Appendix Table~\ref{fig:dns-queries}).  
For example, if an attacker sees a long-running flow to the IP address corresponding to ``dropcam.com,'' that flow is definitely from a Nest camera,
and the traffic pattern of this flow can be used for activity inference. This ability to fingerprint based on destination IP addresses is specific to the IoT setting. Unlike web browsing or other general-purpose Internet traffic, IoT devices typically communicate with only a few servers operated by their manufacturer, and only one flow is typically needed to perform activity inference. 
Destination IP address fingerprinting is also effective even if multiple device signals overlap in time or frequency, because individual packets and corresponding IP headers can still be demultiplexed. We do not consider the potential of background traffic from non-IoT devices to disrupt this fingerprinting process, because we wish to consider a worst case scenario for the defender, and because most web browsing traffic will not involve IoT device cloud servers.

Local adversaries with network access can also identify devices and demultiplex traffic using  MAC addresses and other LAN-available information with a system such as Fingerbank~\cite{fingerbank} or IoT Sentinel~\cite{miettinen2017iot} in addition to destination IP address fingerprints.

If DNS requests or destination IP addresses are obfuscated, such as for an external adversary outside a VPN, for a local adversary without Wi-Fi network access, or through DNS over TLS (DoT)~\cite{RFC-DoT}, DNS over HTTPS (DoH)~\cite{RFC-DoH}, or Oblivious DNS (ODNS)~\cite{schmitt2018oblivious}, then device fingerprinting would have to be performed using traffic rate characteristics~\cite{apthorpe2017closing, shamsi2017faulds}. 
Our STP algorithm (Section~\ref{sec:iilp}) protects against user activity inference from traffic rates independent of the device identification technique employed by the adversary.

\vspace{-18pt}
\subsubsection{Activity Inference from Traffic Rates}
\vspace{-10pt}
\label{sec:case-studies}
Once traffic flows have been associated with their originating devices, the next step in the attack is to use traffic rate changes to infer user activities. We have found that activity inference is effective for all tested devices, but in the interest of space, we provide representative case studies from six devices covering a range of smart home device types:

\textbf{Sense sleep monitor.} Figure~\ref{fig:sleep-pcap} shows send/receive rates from the Sense over a 12 hour period from 10:40pm to 10:40am. Notably, the send/receive rate peaked at times corresponding with user activity. The user shut off the light in the laboratory smart home and went to bed at 12:30am, temporarily got out of bed at 6:30am, and got out of bed in the morning at 9:15am. 
A network observer could already guess when users sleep based on decreases in smartphone or PC web traffic; however, this assumes that smartphone and PC use only stops due to sleep, that everyone in the home sleeps at the same time and does not share devices, and that users do not leave smartphones and PCs to perform network-intensive tasks while they sleep.
The single-purpose nature of the IoT sleep monitor makes none of these assumptions necessary to infer users' sleeping patterns. 

\begin{figure*}[!t]
\centering
\includegraphics[width=0.95\textwidth]{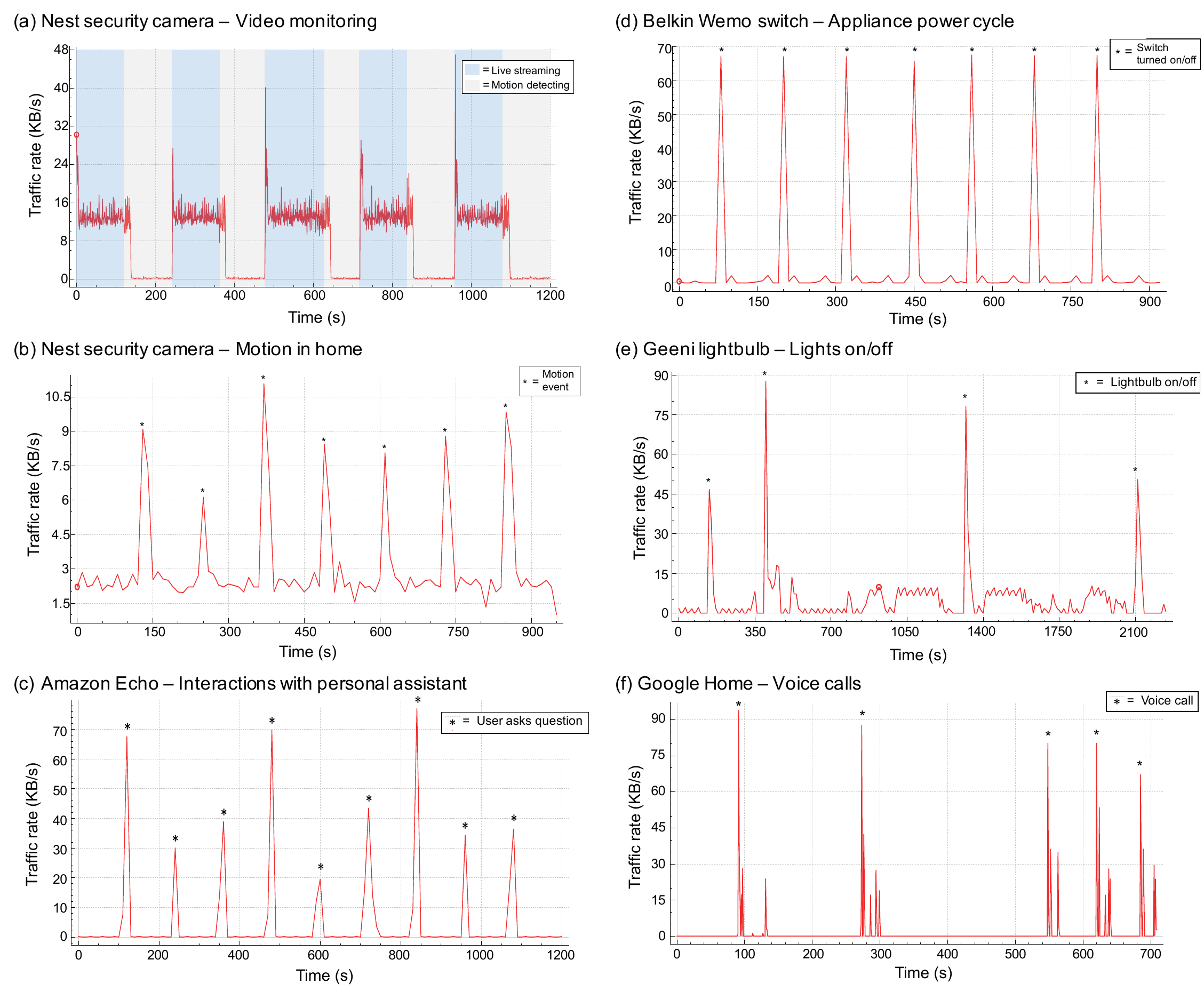}
\caption{Network traffic send/receive rates of selected flows from five commercially-available smart home devices during controlled experiments. Clearly visible changes in traffic rates directly correspond with user activities.  A passive network observer aware of this behavior could easily correlate smart home traffic rates with device states and infer user activities.}
\label{fig:pcap-attack}
\vspace{-10pt}
\end{figure*}

\textbf{Nest security camera.} The Nest Cam Indoor has at least two modes of operation: a live streaming mode and a motion detection mode. In live streaming mode, the camera uploads live video to the cloud for storage and/or real-time viewing on the Nest web or mobile application. 
In motion detection mode, the camera monitors the video feed locally for movement but does not upload video to the cloud. If movement is observed, the camera uploads a snapshot of the video and alerts the user.

Figure~\ref{fig:pcap-attack}(a) shows send/receive rates from the Nest camera alternating between live streaming and motion detection mode every 2 minutes. The traffic rate is orders of magnitude higher in live streaming mode (and a short time afterward until the camera is notified that the user has stopped viewing the stream), allowing an adversary to easily determine whether or not the camera's live feed is being actively viewed or recorded. 

Figure~\ref{fig:pcap-attack}(b) shows that an adversary could easily determine when a Nest camera detects movement while in motion detection mode. The camera was pointed at a white screen with a black square that changed location every two minutes. These simulated motion events triggered clearly observable spikes in network traffic. This predictable variability in network send/receive rates would allow a network observer to infer the presence and frequency of motion inside a smart home.

These issues are significant privacy vulnerabilities and physical security risks. It should not be possible for a third party to determine when a security camera detects movement or is being actively monitored.

\textbf{Amazon Echo.} We tested the Echo by asking a series of 3 questions (``what is the weather?,'' ``what time is it?,'' and ``what is the distance to Paris?'') repeated 3 times, one question every 2 minutes. Figure~\ref{fig:pcap-attack}(c) shows the send/receive rates of SSL traffic between the Echo and a single \texttt{amazon.com} IP address during the experiment. Although the Echo sent and received other TCP traffic to different domains during this time, we were able to identify the stream that correlated with the questions. An adversary could also identify this stream and use the SSL traffic spikes to infer when user interactions occurred.

\textbf{Belkin Wemo switch.} The Wemo switch only has two states, on and off, and its network send/receive rates reflect this duality. Figure~\ref{fig:pcap-attack}(d) shows Wemo network behavior when the switch is turned alternatively on and off every 2 minutes using the Wemo smartphone app and the physical button on the device (both cases result in traffic to the Wemo cloud server). The spike in traffic every time the switch changes state clearly reveals user interactions with the device.

\textbf{Geeni lightbulb.} The Geeni lightbulb also has only on and off states, and these states are reflected in network send/receive rates. Figure~\ref{fig:pcap-attack}(e) shows Geeni lightbulb network behavior when the bulb was turned on and off 4 times over a 37 minute period. The state changes are clearly observable as spikes in traffic rate, which could indicate when someone in a smart home turns the lights on and off. This information could in turn correlate with sleep patterns or home occupancy. 

\textbf{Google Home.} The Google Home is able to make hands-free VoIP calls. We tested the Home by placing 5 calls to different entities over an approximately 13 minute period. Figure~\ref{fig:pcap-attack}(f) shows the send/receive rates of traffic from the Google Home to a single \texttt{*.telephony.goog} domain during the testing period. Spikes in traffic to this domain occurred only during the start of each phone call, making detection of voice calls from the Google Home trivial for any adversary able to identify traffic to this domain, either by observing DNS requests or matching IP addresses.

\vspace{-16pt}
\section{Evaluating Existing Defenses}
\vspace{-4pt}
\label{sec:solution}

In this section, we evaluate three existing techniques known to prevent traffic analysis attacks in other contexts. 
We use two metrics, adversary confidence and bandwidth overhead, to compare the techniques. 
Ultimately, the ratio of adversary confidence to bandwidth overhead determines the effectiveness of the defense technique. Some techniques have a fixed ratio, but others are tunable to users' preferences for privacy versus data usage. 

\textbf{Adversary confidence} is the expected ratio of correct activity inferences to attempted activity inferences by an adversary with no prior knowledge when traffic rate metadata is defended by a particular technique. Lower adversary confidence means that the technique is more effective at protecting user privacy. We define $c_{min}$ as the lowest possible adversary confidence, equivalent to the fraction of time that the user activity occurs. An adversary guessing that activities occur at random times will be correct this fraction of guesses. 
With no defense, adversary confidence $\approx\!1$ for all devices we tested in Section~\ref{sec:attack}. 

Defining adversary confidence assuming no prior knowledge makes the metric generally applicable, because it only incorporates in-band information from network traffic. Adversaries with out-of-band prior probabilities that particular user activities occur at particular times will simply combine these priors with information from traffic analysis to obtain individualized inference confidences. However, since in-band techniques will not affect these prior probabilities, the adversary confidence metric as defined captures the general effectiveness of in-band defenses for comparison. 

Adversary confidence also does not rely on the computational capabilities of the attacker.  This improves the generality of the metric; however, other metrics involving the cost of performing inference in different settings and for different inference algorithms may be worth exploring in future work.

\textbf{Bandwidth overhead} is the ratio of network data sent with and without a given defense technique. For example, a bandwidth overhead of $4$ means that applying the technique results in $4$ times as much traffic (e.g., in bytes) sent on the network than would be sent if the traffic were unprotected. A lower bandwidth overhead is preferable because extra traffic contributes to network congestion and can consume user data caps.

\vspace{-12pt}
\subsection{Firewalling Traffic}
\vspace{-6pt}
\label{sec:blocking}
The simplest technique for preventing activity inference is to prevent an adversary from collecting smart home network traffic in the first place. 
Configuring a firewall to block smart home device traffic is straightforward and would protect devices 
from user activity inference by external adversaries.
However, smart home devices are generally not made to work without an Internet connection, so removing WAN connectivity makes many devices useless.
Even some devices with features that users might expect to involve only local communications, such as turning on an IoT outlet using a smartphone on the LAN, do not function without connections to cloud servers (Appendix Table~\ref{fig:blocking}). 
Additionally, a firewall on the home gateway router would not protect against user activity inference by local adversaries.

\textbf{Adversary confidence}. Firewalling traffic results in an adversary confidence of $c_{min}$ if the adversary is outside the firewall and an adversary confidence of $\approx\!1$ if the adversary is inside the firewall. 

\textbf{Bandwidth overhead}. Firewalling traffic has a bandwidth overhead $<1$, because traffic which would otherwise be sent is prevented from leaving the home. 

 \vspace{-12pt}
\subsection{Virtual Private Networks (VPNs)}
\vspace{-6pt}
\label{sec:vpn}
Another technique for preventing activity inference is to tunnel all smart home traffic through a virtual private network.
A VPN wraps all traffic from an endpoint in an additional transport layer, aggregating it into a single flow with the source and destination IP addresses of the VPN endpoints.
This aggregation could make it difficult to determine which variations in the overall traffic rate observed from outside the VPN correspond to user interactions with individual devices.

However, the effectiveness of a VPN depends on the number of devices behind the VPN, as well as the location of the adversary and the VPN endpoints.  If more devices are behind the VPN, including PCs and smartphones in addition to IoT devices, there may be more total traffic through the VPN tunnel. This may make it more difficult to de-multiplex traffic from individual devices, but the additional protection is inconsistent and difficult to quantify. Individual devices may still have sparse communications and distinctive traffic patterns. Furthermore, if one VPN endpoint is on the home gateway router, then the VPN provides no protection against local adversaries. If the other VPN endpoint is visible to an external adversary (e.g., a server on an ISP's network), then the VPN also provides no protection, because the adversary can simply perform activity inference on traffic after it leaves the VPN. 

We have identified three specific cases where an adversary can infer user activity even if the VPN is optimally located. In each of these cases, an adversary can fingerprint devices using VPN traffic rates alone:

\textbf{1. Single device.} 
If a smart home has only one device, the VPN traffic rate will match that from the device, and the attack can proceed as before.

\textbf{2. Sparse activity.} 
If there are multiple devices that send traffic at different times, 
time periods containing traffic from only a single device would still allow activity inference. 
For example, a smart door lock and smart sleep monitor are less likely to be recording user activities simultaneously. 
Traffic observations from particular times of day are likely to contain non-background traffic from only one of these devices.
This would allow an adversary to identify the active device within a time period and perform activity inference as before. 
Additionally, previous work on Tor website fingerprinting indicates that machine learning techniques could potentially allow adversaries to differentiate traffic from multiple devices active simultaneously~\cite{wang2016realistically}.

\textbf{3. Dominating device.} 
An adversary could also perform activity inference on the device that sends the most traffic if it significantly overshadows traffic from other devices. 
For example, traffic from a security camera uploading live video will dominate traffic from less network-intensive devices, such as smart outlets, making patterns in the camera traffic clearly observable. 

\textbf{Adversary confidence.} As these three cases indicate, the adversary confidence provided by a VPN can range from $c_{min}$ to $1$ depending  on the specific set of devices in the smart home and patterns of individual user's behavior. This adversary confidence variability motivates traffic shaping defense techniques that can guarantee certain levels of privacy protection.

\textbf{Bandwidth overhead}. VPNs have a bandwidth overhead of $\approx\!1$. A small amount of additional traffic is necessary for the creation and maintenance of the VPN tunnel, but this is negligible compared to the amount of traffic from smart home devices.

\begin{table*}[t]
\centering
\begin{tabular}{p{0.26\textwidth}p{0.48\textwidth}p{0.17\textwidth}}
\textbf{Defense} & \textbf{Adversary Confidence} & \textbf{Bandwidth Overhead} \\
\hline
Firewalling Traffic &
	$\approx 1$ inside firewall, $c_{min}$ outside firewall &
	$<1$ 
	 \\

Virtual Private Network (VPN) &
	Varies from $c_{min}$ to $1$ depending on devices \& user behaviors &
	$\approx 1$ 
	 \\

Independent Link Padding (ILP)&
	$c_{min}$  &
	$R_{ILP} / R_{normal}$
	\\

Stochastic Traffic Padding (STP) &
	Tunable from $\approx 1$ to $c_{min}$ with $O(q^{-1})$ &
	Varies with $O(q)$
	\\
\end{tabular}
\vspace{4pt}
\caption{Comparison of defenses against in-home activity inference from smart home device traffic rate metadata.   Adversary confidence and bandwidth overhead are defined in Section~\ref{sec:solution}. $c_{min}$ is the baseline adversary confidence with no defense and is equivalent to the frequency of user activities. $R_{ILP}$ and $R_{normal}$ are the mean traffic rates with and without ILP padding, respectively. $q$ is a parameter of STP that determines the frequency of padding independent of user activities.}
 \vspace{-12pt}
\label{fig:comparison}
\end{table*}

 \vspace{-10pt}
\subsection{Independent Link Padding (ILP)}
\vspace{-6pt}
\label{sec:ilp}

Independent link padding involves shaping upload and download traffic rates to match predetermined rates or schedules, thereby exposing no information about device behavior or user activities to an adversary~\cite{van2015vuvuzela, dyer2012peek, fu2003analytical}. 
Constant rate padding to enforce fixed-size packets with constant interpacket intervals is the simplest form of ILP.
An alternative method is to draw packet sizes and interpacket intervals from probability distributions independent of user behavior.  Performing ILP between individual devices and cloud servers results in the following adversary confidence and bandwidth overhead.

\textbf{Adversary confidence.} ILP provides an optimal adversary confidence of $c_{min}$. No information about user activities is contained in traffic rates after ILP. 

\textbf{Bandwidth overhead.} If the mean traffic rate without ILP is $R_{normal}$, the max traffic rate without ILP is $R_{max}$, and traffic is shaped using ILP to a mean rate $R_{ILP}$, the expected bandwidth overhead will be $R_{ILP}/R_{normal}$. 
However, if $R_{ILP} < R_{max}$, the device will experience network latency due to packet buffering when it attempts to send traffic faster than $R_{ILP}$.  This latency may affect device usability, especially since devices typically require the highest send rate during user interactions. 

Due to this tradeoff between bandwidth overhead and network latency, ILP can efficiently protect two specific classes of smart home devices: 1) devices with relatively constant traffic rates and 2) devices that can tolerate long network latencies. Previous work by Datta, et al. has demonstrated the effectiveness of ILP padding on smart home devices with these properties \cite{datta2018developer}.
The existence of these devices is notable, because ILP padding is usually viewed as too expensive for real-world use~\cite{dyer2012peek}. 

However, many types of smart home devices have both low latency tolerance and traffic rates that spike only during user activities (Figure~\ref{fig:pcap-attack}), resulting in \mbox{$R_{normal} << R_{max}$} and either a high bandwidth overhead or high latency when using ILP. 
To verify this, we tested  the performance of ILP shaping on several devices in our laboratory smart home. We found that performing constant rate ILP on the home gateway router required approximately 40KB/s of overhead traffic in order to provide low enough latency to preserve minimal device functionality~\cite{apthorpe2017spying}. This would result in approximately 104GB of overhead data per month---excessive for smart homes with all but the highest data caps. This amount of extra traffic would also place considerable burden on ISPs if ILP achieved widespread use. 

\vspace{-18pt}
\section{Stochastic Traffic Padding}
\vspace{-6pt}
\label{sec:iilp}

\begin{figure}[t]
\centering
\includegraphics[width=\columnwidth]{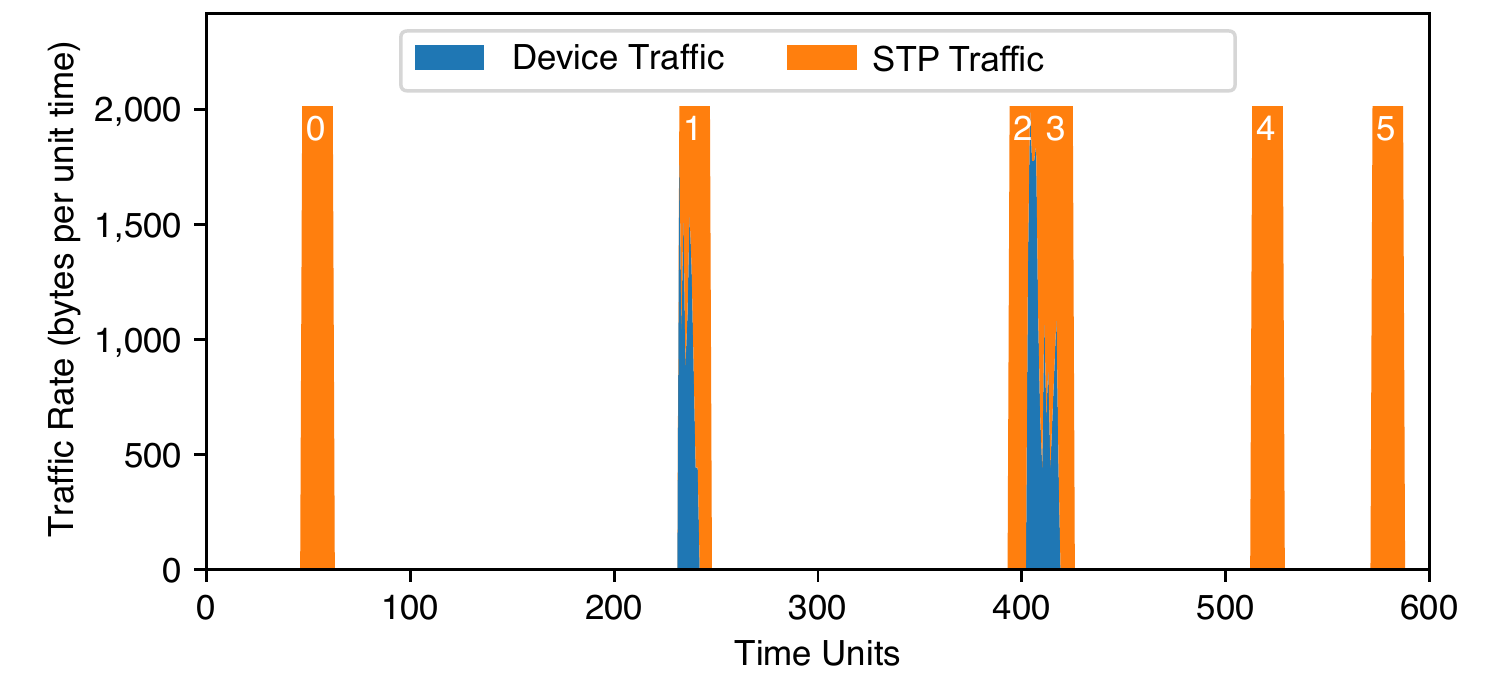}
\caption{Example of STP applied to smart outlet traffic. An adversary could not distinguish which periods of padding mask real device traffic corresponding to user activities.}
\label{fig:simulation-example}
\end{figure}

We introduce stochastic traffic padding (STP), a traffic shaping algorithm to defend against user activity inference from traffic rate metadata. STP provides an easily tunable tradeoff between adversary confidence and bandwidth overhead (defined in Section~\ref{sec:solution}). STP also imposes no additional network latency and can achieve low adversary confidence for relatively little bandwidth overhead. Table~\ref{fig:comparison} compares STP to the techniques discussed in Section~\ref{sec:solution}.
STP can be succinctly described as follows:
\begin{enumerate}
\item Upload and download traffic during user activities is shaped equivalently, so an adversary cannot differentiate different types of user activities (Figure~\ref{fig:simulation-example},  traffic periods 1--3).
\item Additional periods of equivalent shaping are injected randomly into upload and download traffic (Figure~\ref{fig:simulation-example},  traffic periods 0 \& 4--5). An
adversary cannot distinguish these periods from real user activities, reducing confidence in activity inferences.
\end{enumerate}
\vspace{-12pt}
In the following sections, we present the STP algorithm and formalize its adversary confidence and bandwidth overhead (Section~\ref{sec:formal-model}), describe how STP relates to categorical metadata protection (Section~\ref{sec:dns-protect}), evaluate STP using traffic traces from real devices (Section~\ref{sec:simulation}), and discuss how STP can adapt to complicated real-world user behavior (Section~\ref{sec:real-world}).

 \vspace{-12pt}
\subsection{Algorithm and Analysis}
\vspace{-6pt}
\label{sec:formal-model}

\begin{algorithm}[t]
\DontPrintSemicolon
\SetNoFillComment
\small

\SetKwFunction{userActivityOccurring}{userActivityOccurring}
\SetKwFunction{decisionFn}{decisionFn}
\SetKwFunction{uniformRandom}{uniformRandom}
\SetKwFunction{max}{max}
\SetKwFunction{padTraffic}{padTraffic}
\SetKwFunction{STP}{STP}
\SetKwProg{Function}{Function}{:}{}
\SetKwData{padStart}{padStart}
\SetKwData{padEnd}{padEnd}
\SetKwData{padOffset}{padOffset}

\BlankLine

\padStart $\leftarrow 0$\;
\padEnd $\leftarrow 0$\;

\BlankLine

\Function{\STP{$t$, $q$, $T$, $R$}}{
\tcc{Arguments: current time $t$, non-activity padding probability $q$, time period length~$T$, padding rate $R$}

\uIf{$t\!\!\mod T = 0$ \upshape{and \decisionFn{$q$, ...}}}
	{\tcc{ decisionFn() draws a random Boolean from a model parameterized on $q$ and (optionally) previous user activity.}
	\padOffset $\leftarrow$ \uniformRandom{$0$, $T$}\;
	\uIf{$t$ \upshape{+} \padOffset $>$ \padEnd}
		{\padStart $\leftarrow$ $t$ + \padOffset\;
		\padEnd $\leftarrow$ \padStart + $T$\;}
	\uElse
		{\padEnd $\leftarrow$ \padEnd + $T$\;}
	}

\BlankLine
\BlankLine
\uIf{\upshape{\padStart} $\le t \le$ \upshape{\padEnd}}
{\padTraffic{$R$}\;}

\uElseIf{\userActivityOccurring{$t$}}
	{\padStart $\leftarrow$ $t$\;
	\padEnd $\leftarrow$ $t + T$\;
	\padTraffic{R}}
}
\caption{Stochastic Traffic Padding (STP)}
\label{alg:STP}
\end{algorithm}

Algorithm~\ref{alg:STP} presents pseudocode for STP.  
In the following discussion, we refer to traffic corresponding to a privacy sensitive user activity as ``user activity traffic.''

STP begins by choosing a fixed traffic pattern with mean rate $R$ and duration $T$ (Appendix Table~\ref{tab:variables} provides a quick reference for variable definitions used in this section). 
The shape of the traffic pattern is arbitrary, as long as the instantaneous traffic rate across the pattern is high enough that shaping user activity traffic to match the pattern does not impose a latency overhead and the duration of the pattern is longer than the duration of user activity traffic, although this second criterion can be relaxed for a more sophisticated version of the algorithm (Section~\ref{sec:real-world}). 
$R$ can be fixed at the outset if the maximum rate of user activity traffic is known \textit{a priori} or started at a high value and periodically decreased as device traffic is observed.
We use a constant rate traffic pattern for our presentation of STP in order to simplify visualizations, but implementations could also choose a pre-recorded traffic flow scaled to mean rate $R$ or any other predetermined traffic shape.

STP divides time into discrete periods of length~$T$.
All user activity traffic (detected by traffic rate threshold or machine learning methods) is padded to match the preselected traffic pattern, preventing an adversary from differentiating activity types based on traffic rate metadata.  
However, padding user activity traffic alone is insufficient. An adversary would still know that each instance of the fixed traffic pattern corresponds to some user activity, allowing for activity inference from limited-purpose smart home devices (Section~\ref{sec:general-attack}).
STP therefore also performs traffic padding when no user activities occur. 

At the beginning of each period $t$ (such that \mbox{$t \!\!\mod T = 0$}), STP uses a decision function to decide whether to shape traffic during that period. 
If yes, a start time during the period is selected uniformly at random such that $t \le t_{start} < t+T$. Traffic is then shaped from $t_{start}$ to $t_{start} + T$ to match the fixed traffic pattern. 
If shaping is already occurring at $t_{start}$, either due to user activity traffic or padding started during the previous time period, the fixed traffic pattern is simply repeated once the current iteration concludes. This ensures that the total duration of non-interrupted shaping is a multiple of $T$ and that no more than one instance of the fixed traffic pattern starts in each time period.

The following analysis assumes that the decision function performs a random draw from a fixed Bernoulli distribution and that individual user activities occur independently. 
This provides an intuition for the behavior of STP and simplifies derivations of adversary confidence and bandwidth overhead. Sections~\ref{sec:real-world}~\&~\ref{sec:future} discuss ways of extending the decision function and user activity model to handle more complex real-world behavior. 

\textbf{Bidirectional traffic.}
Most smart home devices use bidirectional protocols, especially TCP and HTTP, to communicate with cloud servers.
This means that user activities may be reflected in the patterns of both upload and download traffic. STP must therefore pad traffic in both upload and download directions during user activity or during non-activity periods selected by the decision function.
Provided that $T$ is long enough to cover complete bidirectional communications (requests and responses) corresponding to user activities, all shaped periods will be indistinguishable in both directions.
For example, suppose a device sends a request and receives a response in a single TCP connection with the SYN packet at $t_{SYN}$ and FIN packet at $t_{FIN}$. Bidirectional STP traffic shaping will start at $t_{SYN}$ and continue for $T$, such that $t_{FIN} - t_{SYN} \leq T$. For
DNS and other known UDP protocols, $T$ should
be made long enough
to overlap  the request and response packets
given the latency and bandwidth of the network.

Shaping bidirectional traffic with STP involves two additional considerations. 
First, one direction may have a considerably higher volume of traffic, such as short HTTP GET requests versus longer HTTP responses. 
In this case, using the same fixed traffic pattern in both directions would be wasteful, as one direction requires substantially less cover traffic to mask user activities. 
Instead, STP can use separate fixed traffic patterns with mean rates $R_u$ and $R_d$ for the upload and download directions, respectively. 
Choosing $R_u$ and $R_d$ can be performed as described for $R$ above. 
For the following analyses, we define $R = R_u + R_d$ to reason about the total overhead of STP shaping in both directions.

Second, STP shaping must be applied to upload and download traffic at different locations in the network.
Upload traffic could be shaped on the devices themselves or by a middlebox in the smart home, such as an IoT hub or gateway router. Download traffic could be shaped on the cloud servers or by a middlebox in the network, such as a VPN end point.
These locations must communicate to synchronize periods of padding. This communication could occur through the contents of encrypted cover traffic, but is implementation dependent.
Deciding where to deploy STP determines whether it protects against both internal and external adversaries or external adversaries only (Section~\ref{sec:implementation}).
The following analysis assumes that the adversary can only see traffic shaped by STP.

\textbf{Adversary confidence.}
The adversary's goal is to decide which time periods correspond to user activities.  We can calculate adversary confidence and bandwidth overhead based on the frequency of user activities and the probability of non-activity padding.

We define a probability $p$ that user activity occurs independently during any time period of duration $T$. This probability can be estimated empirically as the fraction of time periods with traffic corresponding to user activities during a representative packet capture. 
We also define a probability $q$ that the decision function chooses to start non-activity padding  independently during any time period. 

Over any given set of $n$ time periods, the expected number with user activities is $np$. However, the expected number of padded periods after STP is $np + n(1-p)q$. Since the adversary cannot tell these periods apart, the expected adversary confidence $c$ is
\begin{equation}\label{eqn:c}
c = \frac{np}{np + n(1-p)q} = \left(1 + \frac{(1-p)q}{p}\right)^{-1}\\
\end{equation}

This is a simple power law that intuitively matches the behavior of STP. 
An adversary continuously performing inference attacks (e.g., an ISP with a permanent tap on a smart home's WAN traffic)
will learn which time periods do \textit{not} contain user activity, but will be unable to determine which of the $c$ fraction of padded periods \textit{do} contain user activity.  
If user activity occurs more frequently (higher~$p$), any particular padded time period is more likely to correspond to a user activity. If non-activity padding occurs more frequently (higher~$q$), any particular padded time period is less likely to correspond to a user activity.

\textbf{Bandwidth overhead.} In order to quantify bandwidth overhead, we define two additional quantities, $D_{A}$  and $D_{\neg A}$, the average amount of bidirectional data sent during periods of user activity and periods of background traffic without user activity, respectively, before~STP.
The expected average bandwidth overhead~$b$ of STP is
\begin{equation}\label{eqn:b}
b = \frac{pRT + (1-p)qRT + (1-p)(1-q)D_{\neg A}}{pD_A + (1-p)D_{\neg A}}
\end{equation}

The values of  $p$, $D_A$, $D_{\neg A}$, and to some extent $R$ are determined by users and device developers, but the STP algorithm can adjust $q$ to trade off adversary confidence and bandwidth overhead. Considering the limits, $q=0$ means that only periods of real user activities are subject to padding, so the adversary can be certain that user activity occurred during these periods. $q=0$ also has the lowest bandwidth overhead, because no cover traffic is sent during periods of no activity.  In comparison, $q=1$ is equivalent to ILP as described in Section~\ref{sec:solution}. $q=1$ has the highest bandwidth overhead because it requires constant cover traffic.

\begin{figure*}[t]
\centering
\vspace{-4pt}
\includegraphics[width=\textwidth]{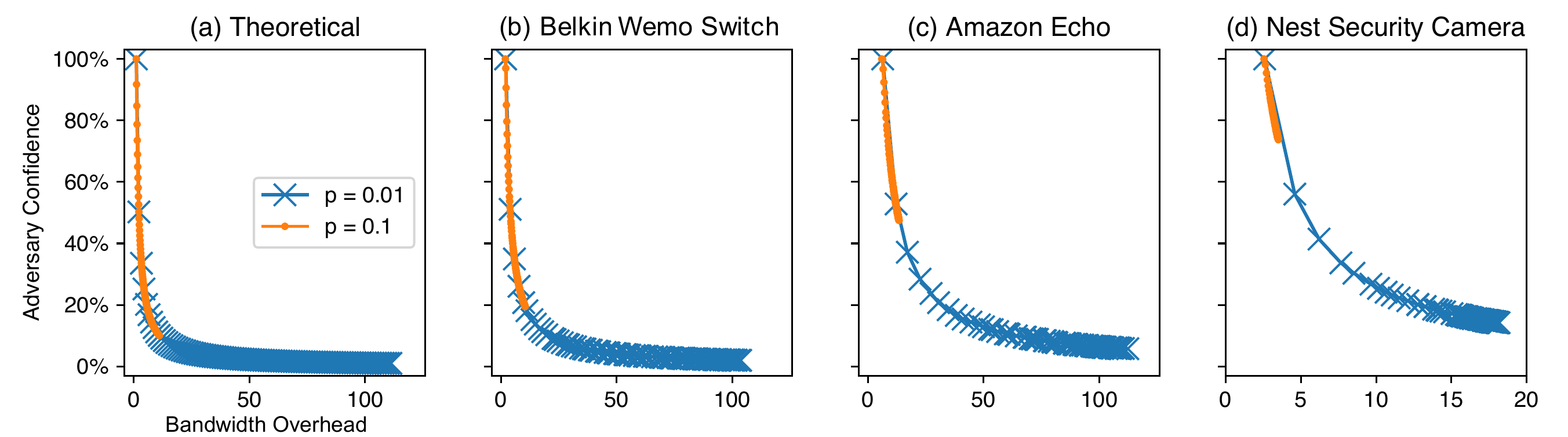}
\caption{STP tradeoff between bandwidth overhead and adversary confidence for different devices and user activity frequencies $p$. Each point corresponds to a probability of non-activity padding $q$ ranging from $0$ (highest adversary confidence) to $1$ (highest bandwidth overhead) in steps of $0.01$.  Note the inverse square relationship, which allows STP to achieve low adversary confidence for relatively little bandwidth overhead.}
 \vspace{-12pt}
\label{fig:simulation-tradeoff}
\end{figure*}

\textbf{Privacy and overhead tradeoff.}
By increasing $q$, adversary confidence decreases according to a power law while bandwidth overhead increases linearly. The ratio of adversary confidence to bandwidth overhead is also a power law:
\begin{equation}\label{eqn:cb}
\frac{c}{b} = O\left(q^{-2}\right)
\end{equation}
To visualize how $c$ and $b$ change as we vary $p$ and $q$, we plot Figure~\ref{fig:simulation-tradeoff}(a) based on Equations~\ref{eqn:c} and \ref{eqn:b}. The horizontal axis shows the bandwidth overhead, $b$, and the vertical axis shows the adversary confidence, $c$. There are two curves, presented in two different colors, which correspond to $p=0.01$ and $p=0.1$, respectively.
Each curve has exactly 101 data points. The leftmost point corresponds to $q=0$. The rightmost point corresponds to $q=1$. We set $RT=1$, $D_A=0.9$, and $D_{\neg A}=0$. These values mean that there is no background traffic, 0.9 units of data are sent per period with user activity, and 1 unit of data is sent per period of padding.

Regardless of $p$, both curves follow a typical power law shape. As $q$ increases from 0, the decrease in $c$ is initially drastic but  flattens as $q$ approaches $1$. This suggests that STP with even a small bandwidth overhead can significantly reduce adversary confidence and improve privacy. However, when $q$ approaches 1, significantly more cover traffic is needed to reduce adversary confidence by the same amount.

It is informative to consider the extreme points on  the curves in Figure~\ref{fig:simulation-tradeoff}(a). The leftmost end where $q=0$ represents the case where padding only occurs during real user activities, allowing an adversary to trivially infer when real user activities take place ($c=100\%$).
In contrast, the rightmost end where $q=1$ and \mbox{$c=c_{min}=p$} represents the case where padding occurs constantly and STP effectively becomes ILP.  Useful settings of $q$ for STP occur between these extremes.

\vspace{-12pt}
\subsection{Obfuscating Categorical Metadata}
\vspace{-6pt}
\label{sec:dns-protect}

STP is focused on protecting \textit{traffic rate} metadata (packet times and sizes) from user activity inference.
However, network traffic flows also contain \textit{categorical} metadata, such as protocols, DNS hostnames, and IP addresses, that could leak information about user activities.
For example, we were able to use DNS hostnames to identify smart home devices (Section~\ref{sec:fingerprinting-demux}).
DNS queries to specific third-party services (e.g., from an Amazon Echo to a music streaming platform~\cite{apthorpe2017smart}) could also directly indicate user activities.

In order to completely protect a smart home from activity inference, STP must be combined with a method to remove or obfuscate categorical metadata related to user activities.
Fortunately, methods to protect categorical metadata of network traffic are widely available.
The cleanest method is to tunnel all STP traffic through a VPN, which groups all smart home traffic (including DNS traffic) into a single flow with a single protocol and packet header information uncorrelated with user behaviors.
Our middlebox STP implementation (Section~\ref{sec:middleboxes}) uses this approach.
While requiring  a VPN raises threshold for adoption, personal VPN usage is becoming more common due to increased privacy awareness~\cite{forbes2018vpn} and availability of non-enterprise VPN services, such as home routers with built-in VPN~\cite{pcmag2018vpn} and Google's Project Fi VPN~\cite{google2018vpn}.
Alternatively, specific categorical metadata could also be protected by existing protocols, such as obfuscating DNS via DNS over TLS (DoT)~\cite{RFC-DoT}, DNS over HTTPS (DoH)~\cite{RFC-DoH}, or Oblivious DNS (ODNS)~\cite{schmitt2018oblivious}. 
DoT, DoH, and ODNS traffic may still need to be shaped separately from non-DNS traffic to prevent web domain inference from packet lengths and other rate characteristics~\cite{RFC-DoT}. STP would then be applied to all traffic to prevent user activity inference.
Because sensitive categorical data can be removed from smart home traffic by existing methods, we focus solely on traffic rate metadata for our presentation of STP.

\vspace{-12pt}
\subsection{Evaluation with Device Traffic}
\vspace{-8pt}
\label{sec:simulation}

The duration and bandwidth of user activity traffic varies across smart home devices and activity types (Figure~\ref{fig:pcap-attack}).  We follow a trace-driven approach to evaluate how adversary confidence $c$ and bandwidth overhead $b$ vary with $p$ and $q$ for STP applied to real device traffic. 

Specifically, we implement device traffic generators that replay traffic recordings from devices in our laboratory smart home. These generators allow us to create realistic traffic traces with varying probabilities $p$ of user activities at different times. We then apply STP to the generated traces at different values of $q$. We find that the relationship between $c$ and $b$ for varying $p$ and $q$ matches the expected power law (Equation~\ref{eqn:cb}) with constant factor variations across devices. 

\textbf{Generating user activity traffic.} We first analyze traffic recordings from three smart home devices (Section~\ref{sec:attack-on-real-devices}) and extract periods corresponding to user activities. These periods include 7 Wemo switch user activities, 9 Amazon Echo activities, and 7 Nest security camera activities. Wemo switch activities last an average of 1 second, Echo activities last 2 to 5 seconds, and Nest camera activities last 8 to 15 seconds.

We then create one generator for each device. At time~$t$, each generator makes a binary decision whether or not to replay an activity with probability $p$. 
If yes, 
the generator randomly chooses a period of recorded user activity traffic and replays it in the generated trace. 
If the replayed traffic lasts $t'$ seconds, the generator will not decide whether to replay another activity until time $t+t'$.
This creates a generated trace with realistic traffic patterns that occur according to a Bernoulli process with tunable probability $p$.
For simplicity, we assume there is no background traffic ($D_{\neg A}=0$).

\textbf{Applying STP.} 
We next apply STP to the generated traffic traces as described in Algorithm~\ref{alg:STP}. We set 
$R$ and $T$ set higher than the maximum bandwidth and duration of any replayed traffic period. 
Figure~\ref{fig:simulation-example} shows an example result from the Wemo switch traffic trace
with $p=0.001$ and $q=0.01$. 
For this particular example,  adversary confidence $ c = 2 / 6 = 33.3\%$ 
and bandwidth overhead $b=6.8$.

\textbf{Visualizing STP trade-offs.} Figure~\ref{fig:simulation-tradeoff}(b--d) shows the trade-offs between $c$  and $b$ for the three device generators with varying $p$ and $q$.
For each device, we vary $p$ from $0.01$ to $0.1$ and increase $q$ from $0$ to $1$ at increments of $0.01$. We run the generators to create traffic traces of over $10,000$ seconds. We create 50 traces per device and plot the mean values of $c$ and $b$.
Similar to the theoretical result in Figure~\ref{fig:simulation-tradeoff}(a), all three curves follow the typical power law shape, which shows increasing bandwidth overhead to achieve the same reduction in adversary confidence as $q$ increases. However, there are three notable differences across devices.

\textit{Bandwidth overhead at $q=0$.} Even at $q=0$ (no non-activity padding), there is still bandwidth overhead due to padding during user activities. For the Wemo switch and the Nest camera, the overhead at $q=0$ and $p=0.01$ is $2.0$ and $2.6$ respectively, while the overhead is 6.4 for the Amazon Echo. This difference is due to the variations in user activity traffic rates across devices. The recorded user activities from the Wemo switch have a standard deviation in traffic rate that is 4.7\% of the mean. This is compared to 28.7\% for Nest camera activities and 59.2\% for Amazon Echo activities.
As $R$ and $T$ are set based on the maximum duration and bandwidth of traffic during user activities for each device, the bandwidth overhead when padding is higher when there are more variations in traffic rates during user activities.

\textit{Curve slope.} The bandwidth overhead required to reduce adversary confidence by the same amount is different across devices. For the Wemo switch, increasing $q$ from $0$ to $0.01$ raises the bandwidth overhead from $2.0$ to $4.0$ and reduces adversary confidence from 100.0\% to 51.1\%. Effectively, $0.04$ average bandwidth overhead is needed per percentage point decrease in adversary confidence.
In contrast, $0.05$ and $0.12$ average bandwidth overhead is needed
for the Nest camera and Amazon Echo, respectively. The bandwidth overhead per unit adversary confidence decrease is highest for the Amazon Echo because its traffic during user activities has the most variations and requires the most cover traffic for constant-rate padding.

\textit{Adversary confidence and bandwidth overhead at $q=1$.} Adversary confidence is highest at $q=1$ for the Nest camera ($c_{min}=14.2\%$). This is because the camera has longer duration user activities that take up a larger fraction of the total time, resulting in less available time for non-activity padding. 
Bandwidth overhead at $q=1$ (when STP effectively becomes ILP) is correspondingly lowest for the Nest camera ($b=18.2$) because traffic during user activities is fairly steady and there is less time to fill with non-activity padding. 

\textbf{Summary.} Using a trace-driven approach, we constructed three traffic generators for the Wemo switch, the Nest camera, and Amazon Echo. We showed that STP bandwidth overhead has an inverse-square relation with adversary confidence for the $p$ and $q$ values we tested. Moreover, we demonstrated that the exact trade-offs between overhead and adversary confidence differ across  devices.
Other devices with similar traffic during user activities as these devices are likely to exhibit similar trade-offs.

 \vspace{-12pt}
\subsection{Adapting to Real-World User Behavior}
\vspace{-6pt}
\label{sec:real-world}

Real-world user behaviors are often more complicated than the simple Bernoulli model assumed in Section~\ref{sec:formal-model}.
While this assumption simplifies activity confidence and bandwidth overhead derivations, it does not limit the generality of STP.
Instead, it highlights where the algorithm could be extended to handle more nuanced user activity patterns that occur in practice.

\textbf{Activity correlations.}
Our derivations of adversary confidence and bandwidth overhead assume that user activities are independent and Bernoulli distributed.
This prevents an attacker from using inter-activity intervals, the amount of time between padded periods, to help distinguish user activities.  For many smart home devices, this assumption holds. Just because a user turns on a lightbulb at a particular time doesn't provide any information \textit{a priori} about when the lightbulb will be turned off. Users may query a personal assistant frequently or infrequently with no apparent pattern. However, other devices, such as a washer with distinct cycles, may have priors on the temporal spacing of user activities that could help an adversary distinguish activity and non-activity shaping.
Packets from unknown bidirectional UDP protocols may also exhibit long-term temporal correlations unknown \textit{a priori.}

\begin{figure*}[t]
\centering
\vspace{-4pt}
\includegraphics[width=0.6\textwidth]{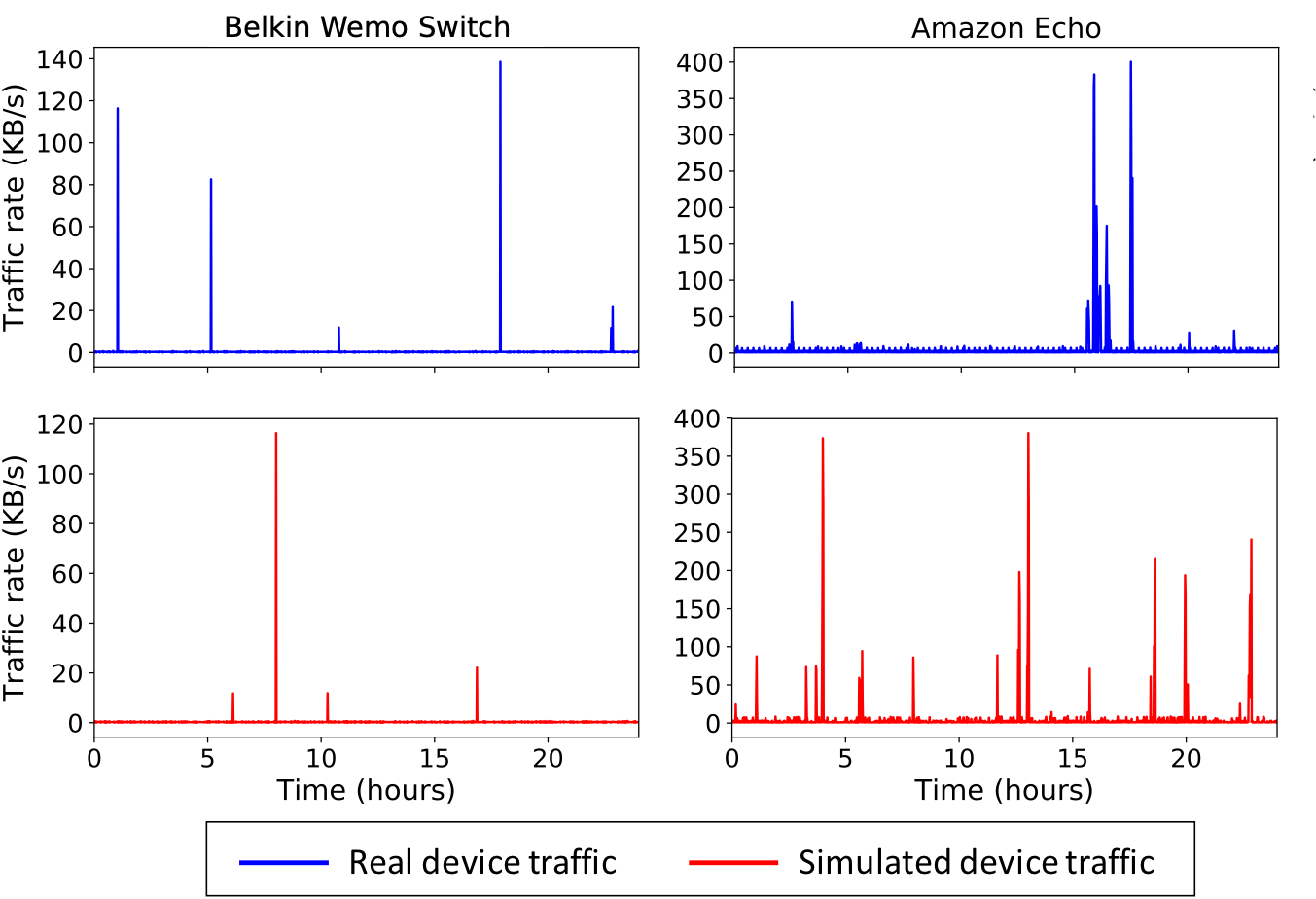}
\caption{The start times of spikes in smart home device traffic traces simulated by hidden Markov models could be used to dictate realistic timings of non-activity padding periods in STP.}
 \vspace{-10pt}
\label{fig:hmm}
\end{figure*}

However, STP is still effective even if user activities exhibit temporal patterns.
The decision function can use a temporal or causal model instead of a Bernoulli distribution to choose when to trigger non-activity padding.
We have found that hidden Markov models can be trained to mimic the patterns of real user activities (Appendix Figure~\ref{fig:hmm}).
Such models could be used to dictate realistic timings of non-activity padding periods. An STP implementation could initially perform ILP (e.g., constant rate padding) while collecting training data from user activity patterns and then switch to STP once a better model has been trained.
This model could then be refined online as more user data is observed.
Of course, even the best models could be fooled by long-term recordings and/or external information, as real user behaviors are driven by variables unavailable to an on-path network device.
Ultimately, STP represents a tradeoff between privacy and overhead traffic volume.

\textbf{Long user activities.}
Some devices may involve user activities of widely variable or unbounded length.
For example, a user watching a live video feed from a security camera may check the feed for a few seconds from a smartphone or may leave the feed open in a browser for an entire afternoon. STP still works in such cases, but we have to relax the assumption that all user activities fit into one time period. Instead, user activities may span multiple time periods, all padded. Non-activity padding must also be allowed to span multiple time periods. The durations of non-activity padding must then be chosen to be statistically indistinguishable from the distribution of real user activity traffic durations. This could be performed by fitting a model to the distribution of user activity traffic durations and using the model to generate non-activity padding durations. 
This model could be continuously refined as more user data is observed.

\textbf{Cross-device correlations.}
The presented STP algorithm also assumes that there are no correlations between user activities across different devices. Such correlations would not be present for non-activity padding periods, making it possible for an adversary to distinguish the times of user activities. For example, consider a home with a smart washer and a smart dryer. It is unlikely that the dryer will be run before the washer. Even if washer and dryer activities separately meet all of the above assumptions, an attacker would still be able to identify some non-activity padding periods by comparing across devices.

Information leakage to external adversaries from cross-device correlations could be prevented by performing STP at the level of an entire smart home instead individually for each device.
Traffic from all devices would be merged into a single flow (e.g. over a VPN from a home gateway router) and then STP could be performed treating the entire home as a single ``device.''
\vspace{-12pt}
\section{STP Implementation}
\vspace{-6pt}
\label{sec:implementation}

In this section, we describe two ways to implement STP: on middleboxes (e.g., home routers) and on IoT devices.

\vspace{-18pt}
\subsection{Implemented on Middleboxes}
\vspace{-6pt}
\label{sec:middleboxes}

We have created a service that enables STP on any Linux-based network middlebox, such as a smart home hub, Wi-Fi access point, or home gateway router. The service
has been tested on the Raspberry Pi Wi-Fi access point in our laboratory smart home and
consists of two components:  a Python script that performs traffic shaping and a custom VPN endpoint.

\begin{figure*}[t]
\centering
\includegraphics[width=0.55\textwidth]{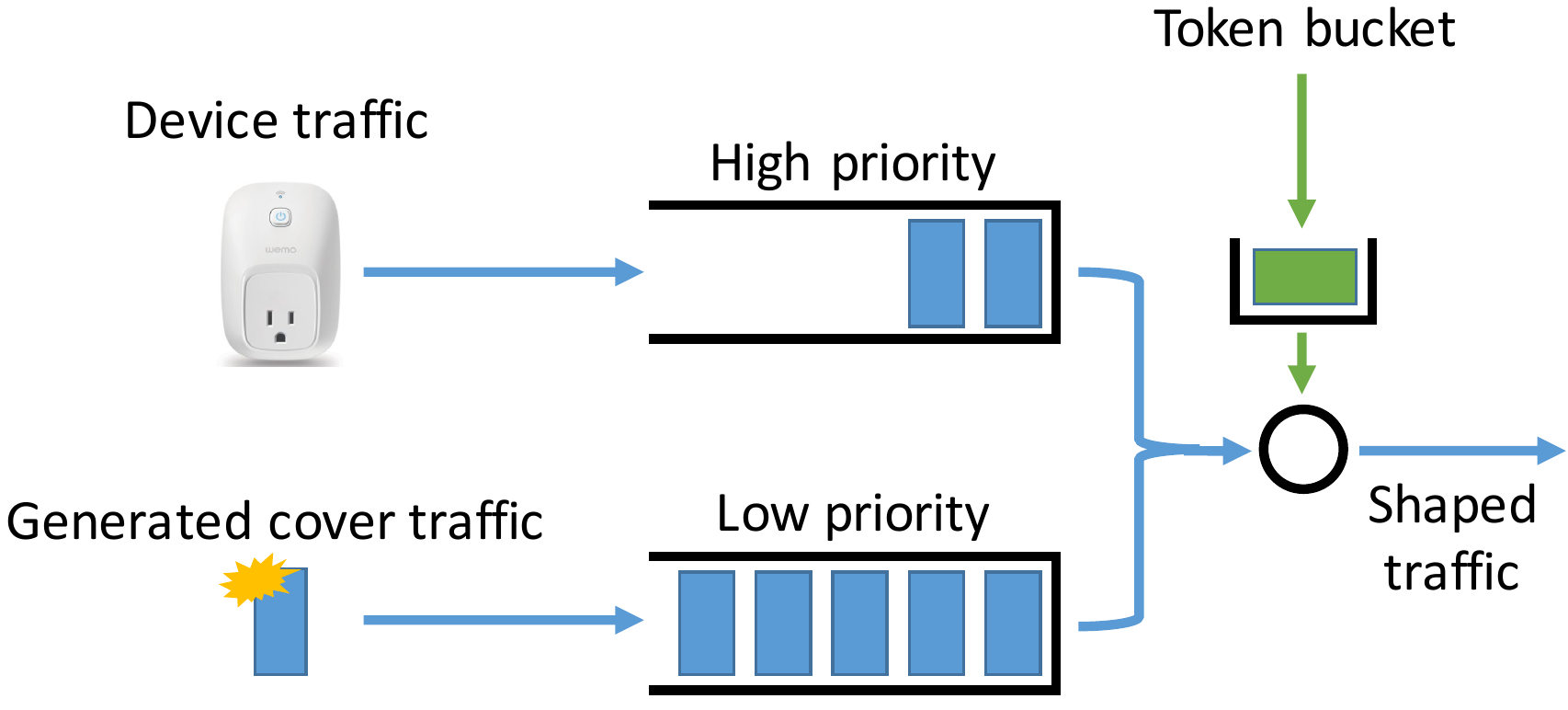}
\caption{Illustration of our traffic shaping implementation for periods of constant rate padding during STP, including cover traffic generation and traffic control in the kernel. Device packets in the high priority queue are always sent before cover packets in the low priority queue. Cover packets are generated faster than the shaped rate, ensuring that cover packets are always present in the low priority queue. The token bucket shaper has a buffer size of 1 token. The rate of token arrival is set to the shaped traffic rate. }
\label{fig:shaper_diagram}
\end{figure*}

\textbf{Traffic shaping.}  The traffic shaping script contains logic to decide when to perform periods of constant rate padding for STP.
The padding itself is implemented using the Linux kernel's traffic control system, configurable via the \texttt{tc} tool, combined with a user-space program that generates cover packets (Appendix Figure~\ref{fig:shaper_diagram}).
The script applies STP to each device behind the middlebox 
(although MAC address filters can be specified to exclude PCs, smartphones, or other non-IoT devices).
The script logic otherwise matches the STP algorithm presented in Section~\ref{sec:formal-model}. Default threshold-based activity detectors and Bernoulli decision functions can be parameterized or replaced if desired.

\textbf{VPN.}
The traffic shaping script automatically connects to an OpenVPN instance hosted on Amazon EC2.
Both VPN endpoints communicate to pad traffic in the upload and download direction, protecting bidirectional protocols as described in Section~\ref{sec:formal-model}. Both VPN endpoints also automatically drop cover packets to prevent them from confusing devices or cloud servers. 
Cover packets are identified as having a destination IP address of the VPN endpoint after exiting the VPN tunnel, while non-cover packets have other destination IP addresses (typically the device, cloud server, or DNS resolver).  In future implementations, any unique flag in cover packet headers or contents could be used to distinguish cover traffic at VPN endpoints. This flag would be encrypted inside the VPN tunnel and indetectable by an adversary with access to the tunneled traffic.

\textbf{Protection.}
Performing STP on a middlebox, such as a  home gateway router, protects against activity inference by external adversaries but not from local adversaries with access to device Wi-Fi traffic.

\begin{figure*}[t]
\centering
\includegraphics[width=\textwidth]{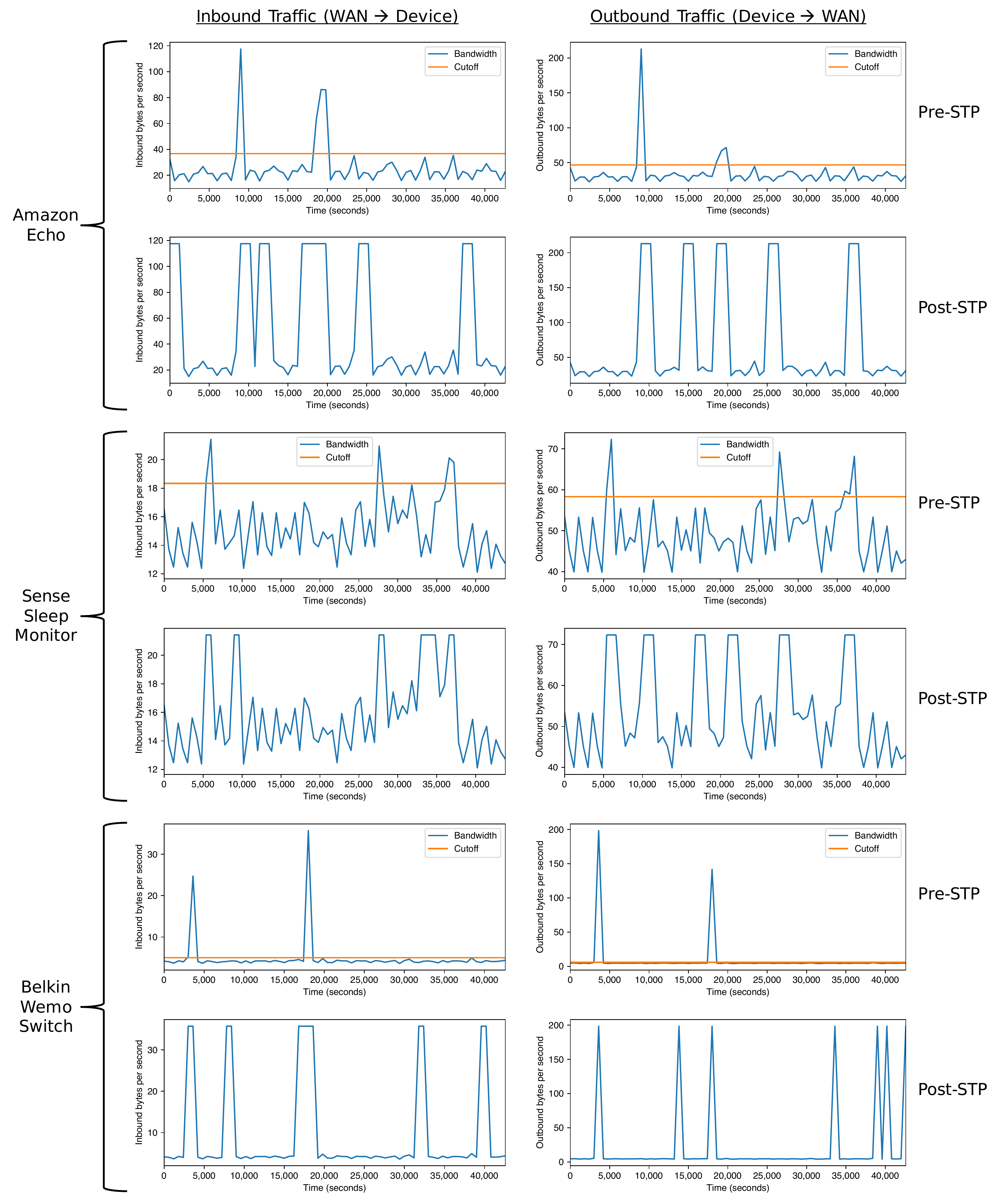}
\caption{12-hour traffic traces from three smart home devices during normal use before and after STP.  STP is applied in both the inbound and outbound directions, with shaped traffic rates $R$ and time period lengths $T$ chosen to cover user activities in each direction. Traffic rate thresholds for triggering padding during user activities are labeled with ``Cutoff'' lines.  For these examples, the probability of injecting periods of cover traffic independent of user activities $q$ was set to $0.05$. As intended, STP adds additional traffic spikes for all devices, reducing adversary confidence in which spikes correspond to real user activities.}
\label{fig:normal-use}
\end{figure*}

\textbf{Empirical bandwidth overhead.}
We replayed 12~hours of Internet traffic from three devices\footnote{Amazon Echo, Sense sleep monitor, Belkin Wemo switch}  in our laboratory smart home through the STP middlebox (Appendix Figure~\ref{fig:normal-use}).  
The traffic was collected during normal use by the smart home inhabitant and contained traffic from 2--3 user interactions per device plus typical background traffic (603KB to 2859KB per device total). Normal device use will vary from home to home, but this analysis demonstrates that the theoretical performance of STP (Section~\ref{sec:iilp}) is observed with real-world device usage patterns.

\begin{figure}[t]
\begin{center}
\includegraphics[width=0.48\textwidth]{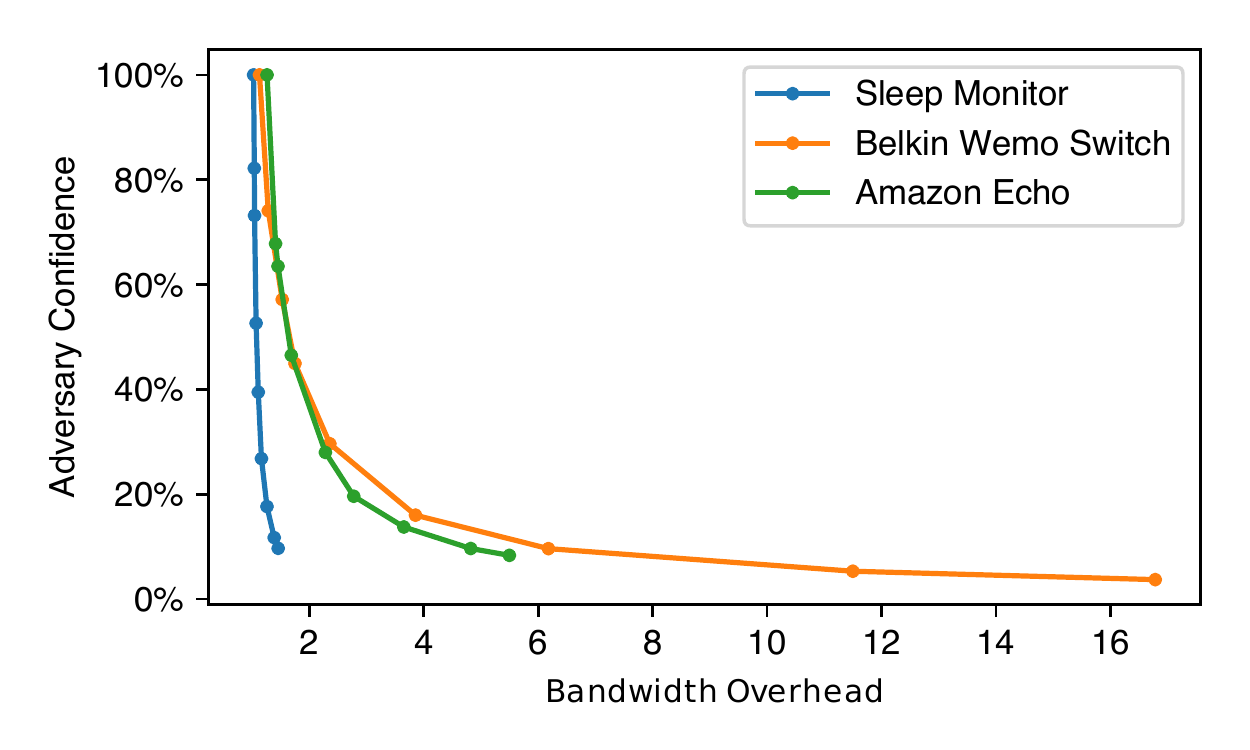}

50\% Adversary Confidence\\
\begin{tabular}{p{2cm}p{2cm} p{2.3cm}}
 & Bandwidth Overhead & Average Absolute Overhead  \\
\hline
Amazon Echo & 1.63 & 3.2 bytes/sec  \\
Sleep Monitor & 1.07 & 0.42 bytes/sec \\
Wemo Switch & 1.68 &  0.80 bytes/sec \\ 
\end{tabular}

\vspace{1.5em}

10\% Adversary Confidence\\
\begin{tabular}{p{2cm}p{2cm} p{2.3cm}}
 & Bandwidth Overhead & Average Absolute Overhead  \\
\hline
Amazon Echo & 4.72  & 19 bytes/sec  \\
Sleep Monitor & 1.44 & 2.4 bytes/sec \\
Wemo Switch & 6.17 & 6.0 bytes/sec \\ 
\end{tabular}
\vspace{4pt}
\caption{STP tradeoff between bandwidth overhead and adversary confidence during 12~hours of real-world use. For all 3 devices, 50\% adversary confidence can be achieved with a bandwidth overhead of 1.7 or less, 
and 10\% adversary confidence can be achieved with a bandwidth overhead of 6.2 or less.}
\label{fig:empirical-overhead}
\end{center}
\vspace{-2em}
\end{figure}

We varied the probability of injecting periods of cover traffic not during user activities, $q$, from $0$ to $1$ to see the tradeoff between bandwidth overhead and adversary confidence (Figure~\ref{fig:empirical-overhead}). 
Running STP to protect a smart home with these devices and similar usage patterns would require only tens of megabytes of extra data per month, a tiny fraction of the tens to hundreds of gigabytes that would be required by constant rate ILP (Section~\ref{sec:ilp}). 
The observed STP bandwidth overhead is also comparable to state-of-the-art padding algorithms from other contexts\footnote{
At 50\% adversary confidence, adaptive padding~\cite{shmatikov2006timing},
Tamaraw~\cite{cai2014systematic}, and WTF-PAD~\cite{juarez2016toward} have reported bandwidth overheads of 
$\approx$2.3, $\approx$1.7, and $\approx$1.2, respectively. 
However, these algorithms do not prevent the user activity inference attack we demonstrate in this paper (Section~\ref{sec:related}).} (Section~\ref{sec:related}).

\vspace{-12pt}
\subsection{Implemented On Devices and Servers}
\vspace{-6pt}
\label{sec:ondevice}
Device developers could alternatively include STP as a feature of their devices, either as custom code or a third-party library.
STP shaping of upload traffic would occur on devices, while shaping of download traffic would occur on cloud servers.
A unique flag in encrypted packet contents (not unencrypted packet headers) would covertly identify cover traffic that should be ignored by devices and cloud servers.
This implementation approach would require the least user effort. It would also allow developers to specify distributions of activity lengths or inter-activity intervals for STP based on device implementation details or the space of possible user interactions with their devices.

Performing traffic shaping on devices would protect first-hop Wi-Fi traffic, preventing both local and external adversaries from performing activity inference. This is important because many users may be more concerned about details of their in-home behaviors leaking to nearby adversaries (nosy neighbors, potential burglers, etc.) than to their ISP.
However, the cover traffic required by STP will place increased burden on device manufacturers' cloud infrastructure. Although this overhead is far less than would be required for ILP shaping,
device developers will likely be unincentivized to include STP on devices without a considerable increase in consumer concern about metadata privacy risks.
In the meantime, traffic shaping at network middleboxes remains the most viable option for privacy conscious smart home device owners.
\vspace{-14pt}
\section{Future Work}
\vspace{-6pt}
\label{sec:future}
Future research efforts could further explore the threat of  activity inference attacks and continue to improve STP and related defense techniques.

\textbf{Fine-grained and higher-order user activity inference.}
The attack we describe involves mostly binary inferences (device state changes) from traffic rate metadata.
We are next interested in whether combining traffic rate metadata with physical layer metadata (e.g., Wi-Fi radio signal strengths) can reveal finer-grained user interactions, such as what smart TV channel a user is watching.
We are also interested in whether combining metadata from multiple devices could allow an adversary to infer higher-order user behaviors, such as ``hosting a party'' or ``late night working.''
Fine-grained and higher-order activity inferences may both constitute privacy violations, and it would be beneficial if researchers could warn consumers about these risks.

\textbf{Individualized adversaries.} Section~\ref{sec:threat} discusses the threat posed by adversaries with prior knowledge about user behaviors.  We assume a low-prior model for this work; however, future research could explore privacy risks of traffic rate metadata against stronger adversaries. Specifically, it would be beneficial to formalize the space of prior knowledge of user activities (ranging from from no prior knowledge to constant physical surveillance) and how the effectiveness of STP and other traffic shaping algorithms vary along this continuum. 

\textbf{Active adversaries.} Future work could also consider user activity inference by active attackers.
Unlike undetectable passive surveillance, active adversaries could interrupt traffic flows and drop packets during periods of padding. 
If this is followed by an unusually high frequency of padded time periods, it may reveal that the dropped packets contained user activity traffic as the device tries to troubleshoot the loss of connectivity. 

\textbf{Improved user interaction models.} 
Our use of hidden Markov models to simulate device behavior (Section~\ref{sec:real-world}) only scratches the surface of future research to improve user activity timing models. Such improved models will be necessary to prevent higher-order correlations from revealing which traffic rate changes in STP indicate real user activities.  However, a strength of STP is that different user interaction models can be easily incorporated to determine the timings of shaped traffic periods. While the specific formulas for adversary confidence and bandwidth overhead will change, the underlying reasoning about the tradeoff between privacy and data use provided by STP will still hold. 

\textbf{Reducing STP bandwidth overhead.}
STP shapes traffic to fixed patterns chosen to cover all possible user activity traffic flows.
This means that relatively low-volume flows from user activities could require large amounts of cover traffic
 to match these fixed patterns.
Future work could further reduce the bandwidth overhead of STP by allowing more fine-tuning of shaped traffic to account for low bandwidth ``mice'' and high-bandwidth ``elephant'' flows.
Rather than $R_u$ and $R_d$, STP implementations could have more options for  shaped traffic patterns with mean rates $[R_0, R_1, \dots]$.
Time periods with real user activity would be padded to the pattern with the minimum $R_x$ that still covers the device traffic.
This is reminiscent of defenses against website fingerprinting proposed by Nithyanand et al.~\cite{nithyanand2014glove} and Wang et al.~\cite{wang2014effective}, which shape finite length traffic flows to match supersequences over anonymity sets of packet sequences.

However, care would have to be taken when choosing which patterns to use for time periods without user activity.
The frequency and timing of each pattern could create a new channel that leaks information about the likelihood of these periods containing only cover traffic.
Combining multiple shaped traffic patterns with the real-world considerations discussed in Section~\ref{sec:real-world} would introduce additional multi-variable relationships to STP, complicating adversary confidence and bandwidth overhead derivations into a topic for future work.
\vspace{-18pt}
\section{Related Work}
\vspace{-6pt}
\label{sec:related}

This paper draws on a rich history of related research on traffic analysis attacks and prevention techniques.  The attack we describe is similar in spirit to the Fingerprint and Timing-based Snooping (FATS) attack presented by Srinivasan et al. in 2008~\cite{srinivasan2008protecting}. The FATS attack involves activity detection, room classification, sensor classification, and activity recognition from Wi-Fi traffic metadata from a sensor network deployed in the home, the precursor to today's smart home IoT devices. In contrast to our attack, FATS relies on 
radio fingerprinting and signal attenuation measurements that are not available to external adversaries. 

Other research has demonstrated traffic analysis attacks on specific IoT devices~\cite{apthorpe2017smart, apthorpe2017closing, grover2016internet}.  Copos et al.~\cite{copos2016anybody} used metadata to detect transitions between Home and Auto Away modes of Nest Thermostat and Nest Protect devices. Our work re-emphasizes metadata privacy concerns demonstrated by these projects for a broader range of modern smart home devices. 

Our attack also draws from side-channel privacy attacks using network traffic metadata on anonymity networks~\cite{back2001traffic, murdoch2005low}, Internet browsing patterns~\cite{felten2000timing, gong2012website}, and user/device fingerprinting \cite{bellovin2002technique, kohno2005remote, verde2014no}.

Similarly, our development of STP was motivated by existing work on traffic shaping for privacy. 
Park et al. have described ``activity cloaking,'' a  technique related to STP which is designed to protect against the FATS attack \cite{park2014energy}. Activity cloaking involves some devices generating fake data to mimic actual private activities. This is similar in motivation to STP, but has several important distinctions. Activity cloaking doesn't shape traffic from real activities, requires participation of many devices (and corresponding adoption by many devices/companies), and
is focused on Wi-Fi eavesdroppers rather than WAN observers.

Liu et al. have described a community-based differential privacy framework to protect smart homes against traffic analysis \cite{liu2018epic}. 
Their approach involves sending traffic between the gateway routers of multiple cooperating smart homes before forwarding it to the Internet. This obfuscates the originating home of the traffic with minimal bandwidth overhead. 
However, this approach could result in long network latencies if the homes are not geographically proximal, and the requirement that multiple homes cooperate raises the bar for adoption. 

Finally, STP was motivated by research on traffic shaping to prevent website fingerprinting and flow correlation in anonymity networks, primarily Tor.  
This includes independent link padding algorithms~\cite{van2015vuvuzela, fu2003analytical}, such as BuFLO~\cite{dyer2012peek}, which force traffic to match a predefined schedule or distribution independent of the unshaped traffic, and 
dependent link padding algorithms, in which unshaped traffic patterns affect the shaped output.
Independent link padding algorithms are effective at preventing user activity inference (Section~\ref{sec:ilp}). 
However,
most recent dependent link padding algorithms cannot protect against user activity inference.

Dependent link padding algorithms include adaptive padding, proposed by Shmatikov and Wang, which forces inter-packet intervals of short-lived web communications through an anonymity network node to match a pre-specified probability distribution~\cite{shmatikov2006timing}. 
Wang et al. also designed an algorithm that uses matched packet schedules to prevent an observer of an anonymity network mix node from pairing incoming flows with outgoing flows~\cite{wang2008dependent}.
In 2016, Juarez et al. applied adaptive padding to prevent Tor website fingerprinting (WTF-PAD)~\cite{juarez2016toward}.
Cai et al.~also presented a defense against Tor website fingerprinting (Tamaraw) that shapes website downloads to multiples of a padding parameter $L$ packets~\cite{cai2014systematic}. 
Wang and Goldberg have also proposed a defense against website fingerprinting (Walkie-Talkie) that uses half-duplex communication to limit the information available to the adversary~\cite{wang2017walkie}.

These dependent link padding techniques allow periods of higher or lower traffic rates to be preserved in the shaped output as long as the traffic is smoothed to be indistinguishable from traces from other websites. 
However, unlike STP, none of these techniques introduce periods of high traffic rates during device (or browser) quiescence to confuse adversaries about when user activities occur. 
Applying these techniques to smart home traffic would still allow an adversary to perform user activity inference, because fluctuations in shaped traffic rates would still be likely correlated with user activities.

Ultimately, we cannot use one of these existing techniques instead of STP, because defending against website fingerprinting is a fundamentally different problem than user activity inference with quite different assumptions and goals.
Website fingerprinting involves comparisons \textit{between} traffic traces (e.g. ``Is this trace similar to previous traces known to be from fetching a particular website?'') while user activity inference involves analysis of patterns \textit{within a single trace} (e.g., ``Is this traffic spike substantively higher than the background, suggesting a user interaction has occurred?''). 
In other words, website fingerprinting asks ``which website is being fetched?'' while user activity inference need merely ask ``is any website being fetched?'' Any use of a single-purpose smart home device may indicate a behavior that a user considers private.
\vspace{-14pt}
\section{Conclusion}
\vspace{-6pt}
\label{sec:conclusion}

The privacy threat from Internet traffic metadata will continue to grow along with the market for IoT smart home devices. 
In this paper, we show that a passive network adversary can infer  private in-home user activities from smart home traffic rates \textit{even when devices use encryption}. 
We introduce ``stochastic traffic padding'' (STP), a traffic shaping algorithm which uses intermittent periods of traffic padding to limit the information revealed about user activities through traffic rate metadata. 
STP provides a tunable tradeoff between adversary confidence and bandwidth overhead, allowing sufficient privacy protection without significantly decreasing network performance or consuming data caps. We demonstrate the effectiveness of STP on traffic traces from real smart home devices and present an implementation for smart home hubs, Wi-Fi access points, and gateway routers. 

\section*{Acknowledgments}
We thank Gunes Acar, Trisha Datta, Frank Li, Srikanth Sundaresan, and the students in the IoT project group of the 2018 Princeton AI4All program:  Vibha Aramreddy, Michaela Guo, Pavitra Kotecha, Dennis Kwarteng, Hung Nguyen, Alberto Olivo, Adithi Raghavan, and William Zafian.  This work was supported by the Department of Defense through
the National Defense Science and Engineering Graduate Fellowship (NDSEG) Program, a Google Faculty Research Award, the Princeton University CITP IoT Consortium,
and NSF Awards CNS-1526353, CNS-1539902, and CNS-1739809.

{\footnotesize \bibliographystyle{acm}
\bibliography{main}}

\clearpage

\begin{table*}[t]
\section*{Appendix}
\begin{center}
\footnotesize
\begin{tabular}{ll}
\textbf{IoT Device} & \textbf{Identifying DNS Query} \\
\hline
Amcrest Security Camera & \url{dh.amcrestsecurity.com} \\
Amazon Echo & \url{device-metrics-us.amazon.com} \\
Belkin Wemo Switch & \url{prod1-fs-xbcs-net-1101221371} \\
D-Link Wi-Fi Camera & \url{signal.auto.mydlink.com} \\
Geeni Lux lightbulb & \url{a.gw.tuyaus.com} \\
Google Home & \url{clients1.google.com} \\
Nest Cam Indoor & \url{nexus.dropcam.com} \\
Orvibo Smart Socket & \url{wiwo.orvibo.com} \\
Phillips Hue Starter Set & \url{diagnostics.meethue.com} \\
Samsung SmartCam & \url{xmpp.samsungsmartcam.com} \\
Samsung SmartThings Hub & \url{dc.connect.smartthings.com} \\
Sense Sleep Monitor & \url{sense-in.hello.is} \\
TP-Link Smart Plug & \url{devs.tplinkcloud.com} \\
Wink Hub & \url{agent-v1-production.wink.com} \\
\end{tabular}
\caption{DNS queries from smart home devices during a representative packet capture that are easily attributable to a specific device or manufacturer. \vspace{0pt}
}
\label{fig:dns-queries}
\end{center}
\end{table*}

\begin{table*}[t]
\vspace{36pt}
\begin{center}
\small
\begin{tabular}{lll}
\textbf{Device} & \textbf{Functionality} & \textbf{Description} \\
\hline
Amazon Echo & limited & Can use as a bluetooth speaker with previously paired smartphone \\
&& Echo recognizes ``Alexa" keyword but does not provide any voice-control features \\ \hline
Belkin Wemo Switch & limited & Can turn switch on/off with physical button on device \\
&& Cannot use smartphone app to control device even when phone on local network\\ \hline
Orvibo Smart Socket & limited & Can turn switch on/off with physical button on device or smartphone app on local network \\ \hline
TP-Link Smart Plug & limited & Can turn switch on/off with physical button on device or smartphone app on local network \\ \hline
Nest Security Camera & none & Unable to view video feed or receive detected motion notifications \\ \hline
Amcrest Security Camera & none & Unable to view video feed or control camera direction\\ \hline
Sense Sleep Monitor & none & Monitor does not record sleep data\\&& Light-based UI does not reflect local sensor readings \\
&& Cannot use smartphone app to control device or access current data \\
\end{tabular}
\caption{Tested commercially-available IoT devices had limited or no functionality when firewalled to prevent communication outside of the smart home LAN.}
\label{fig:blocking}
\end{center}
\end{table*}

\begin{table*}[t]
\centering
\begin{tabular}{cp{6.5cm}}
Variable & Definition \\
\hline
$c$ & Adversary confidence \\
$c_{min}$ & Minimum adversary confidence \\
$b$ & Bandwidth overhead \\
\hline
$p$ & Probability of user activity during any time period \\
$q$ & Probability of padding during time periods independent of user activity\\
\hline
$R$ & Mean rate of all padded traffic ($R_u + R_d$)\\
$R_u$ & Mean rate of padded traffic in upload direction \\
$R_d$ & Mean rate of padded traffic in download direction \\
\hline
$T$ & Time period length \\
$t$ & Time \\
\hline
$D_A$ &  Mean traffic (data) volume during time periods with user activity\\
$D_{\neg A}$ & Mean background traffic volume during time periods without user activity\\
\end{tabular}
\vspace{0.25em}
\caption{Variables used in STP presentation and evaluation. Adversary confidence ($c$) and bandwidth overhead ($b$) are described in Section~\ref{sec:solution}. All other variables are described in Section~\ref{sec:formal-model}.}
\vspace{-1em}
\label{tab:variables}
\end{table*}

\end{document}